\documentclass[sn-mathphys-num,iicol]{sn-jnl}

\usepackage{graphicx}%

\usepackage{multirow}%
\usepackage{amsmath,amssymb,amsfonts}%
\usepackage{amsthm}%
\usepackage{mathrsfs}%
\usepackage[title]{appendix}%
\usepackage{xcolor}%
\usepackage{textcomp}%
\usepackage{manyfoot}%
\usepackage{booktabs}%
\usepackage{algorithm}%
\usepackage{algorithmicx}%
\usepackage{algpseudocode}%
\usepackage{listings}%
\usepackage{ragged2e}
\usepackage{makecell}
\usepackage{subcaption}
\usepackage{placeins}
\usepackage{booktabs}
\usepackage{caption} 
\usepackage{bm}
\usepackage{bbm}
\usepackage{placeins}

\theoremstyle{thmstyleone}%
\theoremstyle{thmstyletwo}%

\theoremstyle{thmstylethree}%

\makeatletter
\renewcommand{\abstractfont}{%
  \reset@font\fontsize{9bp}{11bp}\selectfont
  \normalfont              
  \mathversion{normal}     
  \leftskip=24pt
  \rightskip=24pt
  \parfillskip=0pt plus 1fil
}
\makeatother

\begin{document}

\title[Article Title]{Impact of neutral fluxes and signal significance optimization on semi-exclusive $pp \rightarrow t\bar{t}$ production via deep learning training}


\author[1]{\fnm{A.} \sur{Cota Rodríguez}}
\email{antonio.cota@unison.mx}

\author[2]{\fnm{Jesus Alberto V.} \sur{Corral}}
\email{jesus.corral@ku.edu}
\equalcont{These authors contributed equally to this work.}

\author[1]{\fnm{J.A.} \sur{Murillo Quijada}}
\email{javier.murillo@unison.mx}
\equalcont{These authors contributed equally to this work.}

\affil[1]{\orgdiv{Departamento de Investigaci\'{o}n en F\'{i}sica}, 
\orgname{Universidad de Sonora}, 
\orgaddress{\city{Hermosillo}, \state{Sonora}, \country{México}}}

\affil[2]{\orgdiv{Department of Physics and Astronomy}, 
\orgname{The University of Kansas}, 
\orgaddress{\city{Lawrence}, \state{Kansas}, \country{United States}}}


\abstract{Photon flux benchmark models together with recent experimental estimations for pomeron energy fluxes and structure functions are implemented within Monte Carlo simulation, to determine their impact on physical observables and on the signal strength uncertainty of the yet-to-be observed semi-exclusive $t\bar{t}$ production in proton–proton ($pp$) collisions at Large Hadron Collider (LHC) and Future Circular Collider (FCC) energies.
Expected cross-section rises by a factor of $\sim$50 and $\sim$22 for pomeron-induced and photon-induced processes respectively from LHC to FCC energy regime. TensorFlow deep neural networks were implemented to discriminate semi-exclusive processes against non-peripheral $t\bar{t}X$ background, achieving an Area Under the Curve (AUC) test performance of $0.9917 \pm 0.0002$ for the photon-induced signal. At detector level, following the geometry of the CMS detector at CERN, the minimum Hadronic Forward energy observable was identified as the most effective discriminator against non-peripheral $t\bar{t}X$ background. In the context of low pileup $pp$ data and based on the Asimov dataset, by considering the statistical effects and the systematic contribution from pomeron/photon schemes to the total uncertainty a 5-$\sigma$ significance is expected for pomeron-induced and photon-induced $t\bar{t}$ production modes with 1\,fb$^{-1}$ and 4.7\,fb$^{-1}$ integrated luminosity datasets respectively. The systematic contribution to the total uncertainty is 25.1\% and 7.6\% respectively, highlighting the potential for experimental observation using Run 2 and Run 3 LHC data, allowing further studies within the LHC forward physics program.}

\keywords{Semi-exclusive top quark pair production, photon- and pomeron-induced processes, high energy physics phenomenology, deep neural networks in forward physics, LHC and FCC sensitivity studies.}



\maketitle

\section{Introduction}\label{sec1}


Difractive and photon-induced top–antitop ($t\bar{t}$) production in proton-proton ($pp$) collisions at the Large Hadron Collider (LHC) offers a unique probe of neutral color–singlet exchange mechanisms arising from quantum chromodynamics (QCD) and quantum electrodynamics (QED)~\cite{Howarth:2020uaa, PhysRevD.88.054025, Pitt:2023mtm}. 
In this context, events are categorized according to the dissociation of the beam protons. 
The term \emph{semi-exclusive} refers to configurations in which one proton remains undissociated in the final state, whereas \emph{exclusive} production denotes interactions where both protons emerge undissociated. 
Semi-exclusive processes exhibit distinctive forward–central topologies that contrast with the more isotropic particle production characteristic of non-peripheral interactions~\cite{HarlandLang:2016epjc, atlas_2012_rapidity_gap}, the term peripheral refers to collisions occurring at large impact parameters, where the hadronic overlap between the protons is reduced, while non-peripheral interactions correspond to smaller impact parameters with significant hadronic overlap~\cite{K_usek_Gawenda_2016}. Studies have shown that semi-exclusive processes can occur more frequently than fully exclusive ones by up to three orders of magnitude in certain final states. For $t\bar{t}$ production, while the difference is smaller, it remains significant. Consequently, semi-exclusive production modes emerge as promising candidates for the observation of color–singlet $t\bar{t}$ production beyond the dominant QCD mechanisms of gluon–gluon fusion ($\sim90\%$) and quark–antiquark annihilation ($\sim10\%$)~\cite{ablat2024, Pitt_2024}.\\



\noindent In hadronic collisions, a rapidity gap refers to an extended region in pseudorapidity of suppressed final state particle multiplicity. While smaller gaps can occur in any hadronic collision, large and well-defined gaps are characteristics of processes mediated by neutral color-singlet exchanges, such as photons $(\gamma)$, or pomerons $(\mathbb{P})$. By contrast, non-peripheral production proceeds through color exchange, which has higher particle multiplicity and a more isotropic production of particles. Heavy quark pair production at the LHC can proceed through various partonic interactions~\cite{PhysRevD.93.034038}. Among these, the pomeron–gluon $(\mathbb{P}g)$ and photon–gluon $(\gamma g)$ configurations were regarded as particularly relevant, as they correspond to semi-exclusive topologies characterized by color-singlet exchange and comparatively larger cross sections. These processes offer a framework to explore color–neutral interactions while preserving experimentally accessible event yields. In contrast, the fully exclusive modes $(\gamma\gamma)$, $(\mathbb{P}\mathbb{P})$, and $(\mathbb{P}\gamma)$ are strongly suppressed. The key difference is that exclusive production must proceed entirely through the color–singlet exchange with less QCD activity, while semi-exclusive production allows limited proton dissociation and therefore accesses a much larger phase space~\cite{Howarth:2020uaa}.\\

\begin{figure}[!h]
    \centering
    \begin{subfigure}[b]{\columnwidth}
        \centering
        \includegraphics[width=\columnwidth]{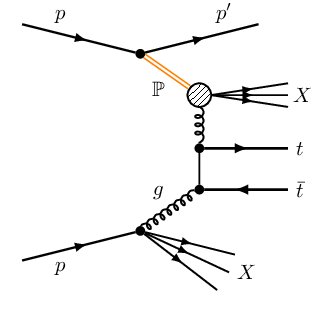}
        \caption{}
        \label{fig:ttbar-pomeron}
    \end{subfigure}

    \vspace{1.0cm} 

    \begin{subfigure}[b]{\columnwidth}
        \centering
        \includegraphics[width=\columnwidth]{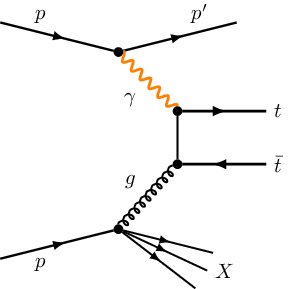}
        \caption{}
        \label{fig:ttbar-photon}
    \end{subfigure}
    
    \caption{\label{fig:diagrams_photon_pomeron} 
    \justifying Representative Feynman diagrams for semi-exclusive top quark pair production: (a) pomeron-induced and (b) photon-induced interactions. In both cases, a color-singlet object is exchanged producing a $t\bar{t}$ pair together with a forward-scattered proton and remnant system $X$.}
\end{figure}

\noindent A schematic representation of the production modes under consideration is shown in Figure~\ref{fig:diagrams_photon_pomeron}, which displays Born-level Feynman diagrams for $t\bar{t}$ production via color-singlet exchange mechanisms. Panel (a) shows pomeron-gluon interactions, $\mathbb{P}g \rightarrow t\bar{t}$, while panel (b) corresponds to photon-gluon interactions, $\gamma g \rightarrow t\bar{t}$. Both processes are mediated by neutral color-singlet exchanges and lead to semi-exclusive topologies. The corresponding semi-exclusive channels can be expressed as: 

\begin{equation}
\text{pomeron-induced:}\quad
pp \to p'\,\bm{\mathbbm{P}}\,\bm{p} \to p'\,\bm{t \bar t\,X}
\label{eq:pomeron_induced}
\end{equation}

\begin{equation}
\text{photon-induced:}\quad
pp \to p'\,\bm{\gamma}\,\bm{p}\to p'\,\bm{t\bar t\,X}
\label{eq:gamma_induced}
\end{equation}

\vspace{0.2cm}

\begin{figure}[h!]
    \centering
    \includegraphics[width=\columnwidth]{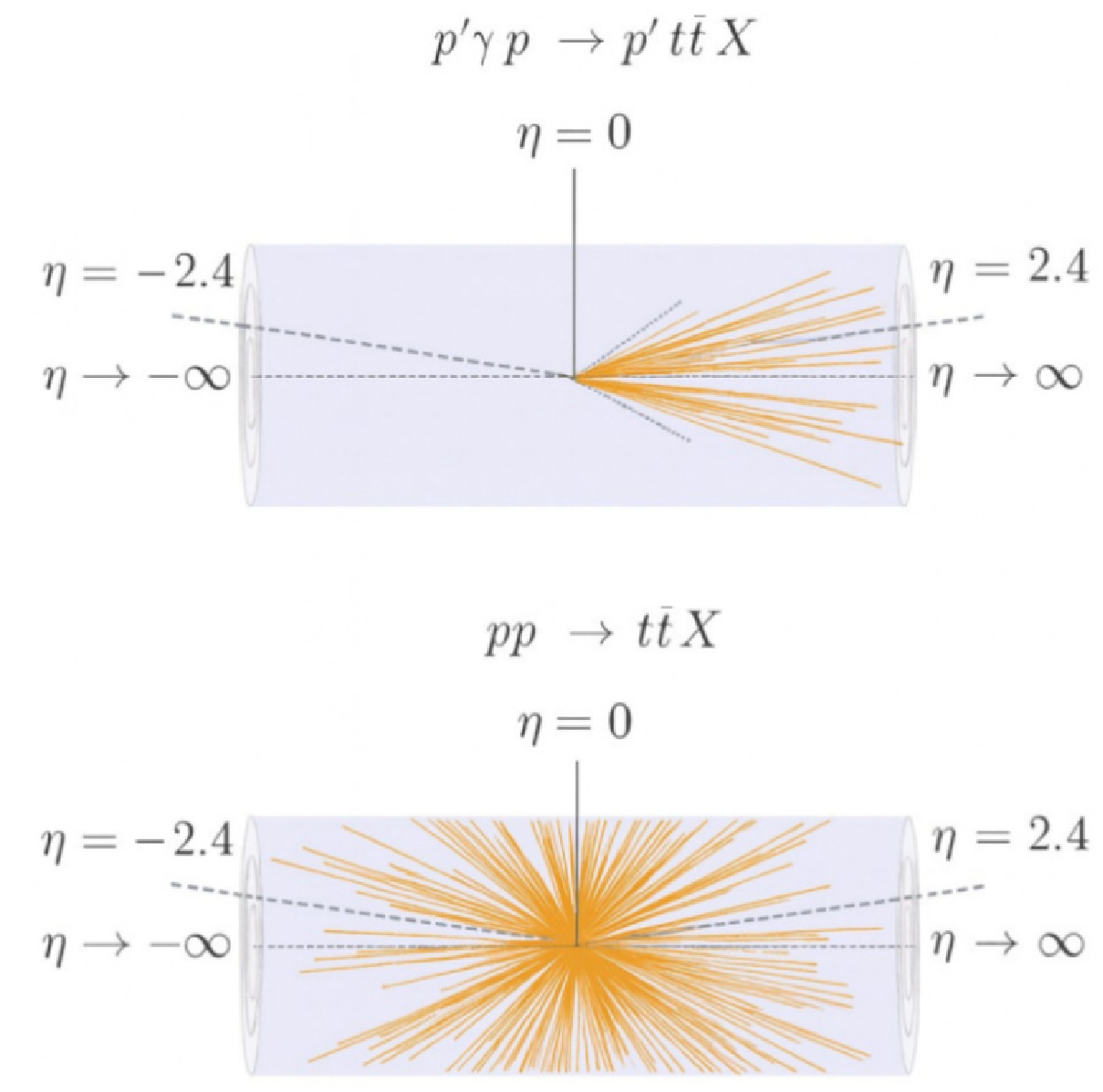}
    \caption{\label{fig:rapidity_gap_illustration}
    \justifying
    Schematic illustration of expected event topologies. The top panel shows a semi-exclusive configuration, such as $p'\gamma p \to p' t\bar{t}X$ or $p'\mathbb{P}p \to p' t\bar{t}X$, characterized by a forward rapidity gap and reduced hadronic activity on one side of the detector. The bottom panel shows a visualization of non-peripheral $pp \to t\bar{t}X$ production.}
\end{figure}

\noindent In the following, we refer to these processes as \textit{pomeron-induced} and \textit{photon-induced} events respectively, according to the nature of the exchanged object involved in the interaction. The expressions in Eqs.~(\ref{eq:pomeron_induced},~\ref{eq:gamma_induced}) use boldface to highlight either the pomeron or photon object with $p'$ being the scattered non-dissociating proton that acts as their source and $p$ the interacting proton that breaks apart. In the right the $t\bar{t}$ system is also highlighted along with additional products from multiple parton interactions and radiation that are represented with the $X$ symbol. Another schematic illustration of the expected event interactions is shown in Fig.~\ref{fig:rapidity_gap_illustration}. The upper panel corresponds to a semi-exclusive configuration, as mentioned before, this topology is characterised by the presence of a rapidity gap, together with reduced particle multiplicity on one side of the detector. When the undissociated proton carries a longitudinal momentum in the positive $z$ direction $(p_{z} > 0)$, the final state particles tends to be boosted such that the majority of produced particle populate the positive pseudorapidity region, leading to the formation of large rapidity gap on the opposite side. By contrast, the lower panel shows non-peripheral $p p \rightarrow t\bar{t}X$ production, where color exchange leads to higher particle multiplicity and more symmetric distribution of final-state 
particles across pseudorapidity range. The size of the rapidity gap, usually denoted as $\Delta \eta$, therefore provides a convenient observable to distinguish semi-exclusive events from such non-peripheral collisions~\cite{atlas_2012_rapidity_gap, nurse2015}. \\

\noindent The top quark, with a mass of approximately 172~GeV, is the heaviest known elementary particle in the Standard Model (SM)~\cite{PhysRevD.110.030001_TopQuark}. Its distinctive properties include the largest Yukawa coupling to the Higgs boson and an extremely short lifetime, decaying in less than $5\times 10^{-25}$~s, well before the hadronization timescale. Due to its large mass value, its production from initial state elements with constrained energy such as photons or pomerons is very reduced in comparison with its production from parton elements from two dissociating protons. No evidence of such $t\bar{t}$ production modes via color-singlet exchange has been observed by LHC experiments so far. These production mechanisms have gained attention as a promising probe to test SM predictions on yet unobserved phenomena and to explore the internal partonic structure of the proton in non-perturbative regimes~\cite{Pitt:2023mtm,Pitt_2024}. Exclusive and semi-exclusive top pair production in $\gamma\gamma$, $\gamma\mathbb{P}$, $\mathbb{P}\mathbb{P}$, $\mathbb{P}p$ and $\gamma p$  configurations have shown to be sensitive to proton structure, pileup conditions, and the capabilities of forward proton tagging~\cite{Martins:2022dfg}. This context motivates a clear characterisation of pomeron- and photon-induced signatures relative to $pp\to t\bar{t}X$ background. Additionally, complementary studies at future electron--proton facilities, such as the LHeC and FCC-he, have demonstrated that tagged top-pair photoproduction can offer precise sensitivity to top quark dipole moments and constrain possible deviations in the $tbW$ coupling~\cite{Bouzas:2021gwx}. The first dedicated search for exclusive $t\bar{t}$ production in proton-tagged events was conducted by the CMS and TOTEM collaborations at $\sqrt{s} = 13~\text{TeV}$, setting an upper limit at 95\% confidence level~\cite{CMS:2023naq}. These previous analyses established feasibility for subsequent studies of top quark pair production via photons and pomeron interactions. \\



\noindent On the experimental side, both major detectors at the LHC have dedicated forward proton spectrometers for tagging intact protons. The CMS–TOTEM Precision Proton Spectrometer (CT–PPS) incorporates silicon tracking and precision timing detectors installed in ``Roman Pots'' stations located approximately 203--219~m from the interaction point, enabling measurements of the proton fractional momentum loss ($\xi$) and squared momentum transfer ($t$). Similarly, the ATLAS Forward Detectors, composed of the AFP and ALFA subsystems, perform analogous measurements for protons scattered at small angles, allowing studies of both exclusive and semi-exclusive processes~\cite{Tumasyan_2023IOP, TRZEBINSKI2023168187}. Additionally, forward calorimetry systems such as the CMS Hadronic Forward calorimeter (HF), covering $2.9 \leq |\eta| \leq 5.2$, and the ATLAS Forward Calorimeter (FCal), covering $3.1 \leq |\eta| \leq 4.9$, provide an alternative means to monitor low energy activity produced by non-dissociating hadrons~\cite{Merlo:1998au, Gulmez:2017zaa, CMS:2024ayq}. Such detectors have allowed the evidence of several photo-induced processes within $pp$, proton-nucleus and ion-ion collisions, for example the case of $pp$ collisions, the observation of exclusive dileptons production $\gamma \gamma \rightarrow l^{+}l^{-}$ via photon-photon fusion ~\cite{CMS:2018uvs, CMS_Collaboration_2024}, together with measurements of single diffractive and double-pomeron exchange processes~\cite{CMS:2015inp}, single-diffractive dijet production and central exclusive hadron pair production~\cite{Sirunyan_2020, PhysRevD.109.112013} within $pp$ collisions has been recently reported. For the case of heavy ion collisions, the first observation of coherent $\phi$ meson production via its kaon decay $K^{+}K^{-}$ channel~\cite{cmscollaboration2025}, and evidence for light-by-light scattering have been achieved~\cite{2019134826}, enabling searches for axion-like particles. The recent measurement of coherent $J/\psi$ photoproduction on the other hand provides new insight into QCD dynamics in nuclear environments~\cite{ATLAS:2025uxr}. These results collectively highlight the growing experimental capability to isolate and study diffractive and photon-induced interactions, reinforcing their importance within the broader LHC forward physics program \cite{Adhikary_2025, Sen:2017dpl, Viti:2011zz}.\\



\noindent This letter focuses on the semi-exclusive $t\bar{t}$ production modes $\mathbb{P}p$ and $\gamma p$ within $pp$ collisions at LHC energies, examining how the predicted $t\bar{t}$ production rates and kinematic observables depend on the modeling of photon and pomeron fluxes and structure functions. Section~2 introduces the phenomenological framework used for pomeron and photo-induced $t\bar{t}$ production. Section~3 describes the corresponding energy fractions and partonic structure of the photon, pomeron, and proton. Section~4 discusses rapidity-gap formation and charged-particle multiplicity at generator level, while Section~5 presents detector-level results and selection efficiencies obtained after full event reconstruction. Section~6 outlines the implementation of the neural-network classifier used to distinguish semi-exclusive processes from non-peripheral $t\bar{t}X$ background. Section~7 reports the likelihood-scan procedure used for signal-strength estimation. Finally, Section~8 summarizes the main conclusions and provides an outlook for future experimental searches or measurements. Such properties and features from generator to detector level will serve as a guide for optimal search on these yet to be discovered semi-exclusive top quark production modes.\\


\section{\label{sec:level2}Pomeron/photon induced $t\bar{t}$ production}


\noindent In diffractive scattering, the exchanged object is commonly modeled by the pomeron $(\mathbb{P})$, a color-singlet Regge trajectory carrying the quantum numbers of the vacuum. Semi-exclusive diffractive production is thus described non-perturbatively through pomeron exchange within the Regge theory~\cite{levin1995}.  In QCD, this picture is formalized through the \textit{diffractive parton distribution function} (DPDF) definition~\cite{Collins:1997sr, PhysRevD.53.6162,PhysRevD.51.3182}, where the pomeron is treated as a virtual exchange object endowed with a partonic structure, analogous to the parton distribution function (PDF) of the proton.  The partonic content of the pomeron, determined in deep inelastic scattering (DIS) experiments~\cite{PhysRevD.110.030001,BARTELS1990175,Salajegheh:2023jgi}, are incorporated into quantum field theory (QFT) calculations through the factorization theorem~\cite{Stewart:2009yx}. In this formalism, the SM production cross section is obtained by integrating over the initial-state partonic momentum fractions weighted by their corresponding PDFs:

\begin{align}
d\sigma &= 
\sum_{i,j}
\int dx_1\,dx_2\,
f_{i/p}(x_1,Q^2)\,
f_{j/p}(x_2,Q^2)\,
\nonumber \\[4pt]
&\times\,
d\hat{\sigma}_{ij\rightarrow X}(x_1,x_2,Q^2)
\label{eq:sigma_pp}
\end{align}

\noindent where $d\hat{\sigma}_{ij\rightarrow X}$ is the partonic hard-scattering cross section. Extending this framework to diffractive scattering, one replaces one of the proton PDFs by the DPDF associated with the pomeron, leading to Eq.~(\ref{eq:sigma_diffractive}).

\begin{align}
d\sigma^{D} &=
\sum_{i,j}
\int dx_1\,dx_2\,
f^{D}_{i/p}(x_1,Q^2;\xi,t)\,
f_{j/p}(x_2,Q^2)\,
\nonumber \\[4pt]
&\times\,
d\hat{\sigma}_{ij\rightarrow X}(x_1,x_2,Q^2),
\label{eq:sigma_diffractive}
\end{align}

\noindent Where $f^{D}_{i/p}(x,Q^2;\xi,t)$ represents the probability of finding a parton $i$ in the proton under the requirement that diffraction occurs with fractional momentum loss $\xi$ and squared momentum transfer $t$.  Within the Ingelman--Schlein (IS) model~\cite{Ingelman:1984ns}, the DPDF is further factorized into a pomeron flux and a parton distribution inside the pomeron, as shown in Eq.~(\ref{eq:IS_factorization}).

\begin{equation}
\underbrace{f^{D}_{i/p}(x, Q^{2}; \xi, t)}_{\text{DPDF}} \;=\;
\underbrace{f_{\mathbb{P}/p}(\xi, t)}_{\text{pomeron flux}} \times
\underbrace{f_{i/\mathbb{P}}(\beta, Q^{2})}_{\substack{\text{parton} \\ \text{distribution} \\ \text{of pomeron}}}
\label{eq:IS_factorization}
\end{equation}

\noindent In this formulation, semi-exclusive $t\bar{t}$ production becomes directly sensitive to the pomeron’s partonic structure, providing a unique probe of its underlying QCD dynamics~\cite{Royon:2022mpk}. Here $f^{D}_{i/p}(x, Q^{2}; \xi, t)$ denotes the DPDF, i.e.\ the probability density of finding a parton $i$ in the proton under the condition that diffraction occurs. On the right-hand side, $f_{\mathbb{P}/p}(\xi, t)$ is the pomeron flux factor describing the probability of the proton emitting a pomeron with fractional momentum loss $\xi$ and squared momentum transfer $t$, while $f_{i/\mathbb{P}}(\beta, Q^{2})$ is the parton distribution inside the pomeron, with $\beta = x/\xi $ the fraction of the pomeron’s momentum carried by the parton and $x$ is the Bjorken variable. In this way, the DPDF is written as the product of the emission probability of the pomeron and its internal partonic structure. The IS factorization introduces a model-dependent uncertainty, $\delta_{\text{model}}$, in addition to the statistical ($\delta_{\text{stat}}$) and systematic ($\delta_{\text{syst}}$) uncertainties:

\begin{equation}
\sigma^{\text{pred}} = \sigma^{\text{true}} 
+ \delta_{\text{stat}} + \delta_{\text{syst}} + \delta_{\text{model}}.    
\end{equation}


\noindent The $\delta_{\text{model}}$ term represents the uncontrolled theoretical uncertainty 
arising from the phenomenological treatment of the pomeron flux and its internal 
parton distributions. Future developments in diffractive factorization could reduce or redefine this term. The applicability of this factorized ansatz was known to be process-dependent. At HERA, DIS measurements showed that the IS--DPDF picture does not hold in \(ep\) collisions~\cite{lee2025}. In $p\bar{p}$ collisions this breakdown is also well supported by data, as Tevatron measurements showed that single-diffractive dijet and electroweak-boson cross sections were overestimated by nearly an order of magnitude when HERA-based DPDFs were used, providing clear evidence for a failure of hard-scattering factorization in that environment~\cite{Akiba_2016}. This suppression may arise from additional spectator interactions that fill the rapidity gap, and can be modelled through a gap--survival probability \(S^{2}\). At the LHC, direct extrapolations of HERA DPDFs were found to over-estimate diffractive cross sections, and phenomenological corrections based on multiple parton interactions (MPI) or gap-survival factors were introduced to reduce these predictions. In simulations, this behavior emerged through the MPI configuration, since samples produced without MPI typically increased the diffractive yield, while those produced with MPI gave estimates more in line with expectations~\cite{Helenius_2019, helenius2021photoproduction}.\\


\noindent In practice, the diffractive cross section was obtained by integrating the convolution of the DPDF and the proton PDF with the partonic hard-scattering cross section $\hat{\sigma}_{ij \to t\bar{t}}$. Although this framework has historically been regarded as a phenomenological ansatz rather than a rigorous consequence of QCD factorization~\cite{Frankfurt_2022}, it provides a good starting point for modeling diffractive processes. Our baseline configuration adopted for single-diffractive $t\bar{t}$ production followed the default \textsc{Pythia}~8.312 setup~\cite{bierlich2022}, which employed the Schuler--Sjöstrand parameterization of the pomeron flux~\cite{PhysRevD.49.2257} together with the H1~2006 Fit~B~LO diffractive PDFs~\cite{aktas_2006}, evolved under the DGLAP equations. 
The proton side was described using the \texttt{NNPDF31\_NLO\_0118\_luxqed} set, provided through the \textsc{LHAPDF} library~\cite{LHAPDF6}. The total cross section for the pomeron-induced channel was evaluated using a factorized approach inspired by the resolved-pomeron model, with MPI checked as defined in Eq.~(\ref{eq:sigma_diffractive_pomeron}).

\begin{multline}
\sigma_{p'\mathbb{P}p} =
S^{2} 
\sum_{i,j} 
\int dx_{\mathbb{P}} \, dt \, d\beta \, dx \;
f_{\mathbb{P}/p}(x_{\mathbb{P}}, t)
\\[4pt]
\times 
f_{i/\mathbb{P}}(\beta, Q^{2}) \,
f_{j/p}(x, Q^{2}) \,
\hat{\sigma}_{ij \rightarrow t\bar{t}}(\hat{s}) ,
\label{eq:sigma_diffractive_pomeron}
\end{multline}


\noindent In the context of semi-exclusive photon-induced production, the quasi-real photon $(\gamma)$ was associated with the electromagnetic field of the non-dissociating proton, and its energy spectrum was described within the Equivalent Photon Approximation (EPA)~\cite{Goncalves_2020}. 
This formalism has been extensively validated in recent LHC observations, including light-by-light scattering~\cite{ATLAS:2017fur} and the semi-exclusive and exclusive production of lepton pairs~\cite{CMS:2018uvs}. Within a similar factorization framework to that employed for the pomeron-induced process in Eq.~(\ref{eq:sigma_diffractive}), the total $t\bar{t}$ production rate in the photon-induced case was obtained by convoluting the EPA photon flux with the proton PDFs and the hard $\gamma p$ scattering cross section, expressed as  Eq.~(\ref{eq:sigma_photon}).

\begin{equation} 
\begin{split} \sigma_{p'\gamma p} = \left. \int d\omega \, \frac{dN}{d\omega} \right|_{p_1} \hat{\sigma}_{\gamma p_2 \rightarrow t\bar{t}}(W_{\gamma p_2}) \\ + \left. \int d\omega \, \frac{dN}{d\omega} \right|_{p_2} \hat{\sigma}_{\gamma p_1 \rightarrow t\bar{t}}(W_{\gamma p_1}) 
\label{eq:sigma_photon}
\end{split} 
\end{equation}

\vspace{0.25cm}

\noindent where \( \omega \) is the photon energy, \( \frac{dN}{d\omega} \) is the photon flux considering two possible directions for non-dissociating proton 1 and 2 (right or left), $\hat{\sigma}_{\gamma p \rightarrow t \bar{t}}$ denotes the $\gamma p$ partonic cross section already integrating over Bjorken momentum fractions. The photon flux is modelled via EPA, using the Weizsäcker-Williams method for virtual photons \cite{Goncalves2013}. Events were simulated at generator level using \textsc{MadGraph5\_aMC@NLO}~2.8.3.2~\cite{Alwall_2011} with a Next-to-Leading Order (NLO) precision level, and subsequently passed to \textsc{PYTHIA}~8.312 for showering and hadronization stages. To evaluate the dependence of the photon and pomeron flux and structure scheme model over the total cross section, additional benchmark schemes were simulated alongside the default setup indicated above for both the \(p'\mathbb{P}p \rightarrow p't\bar{t}X \) and \(p' \gamma p \rightarrow p't\bar{t}X\) production mechanisms. Table~\ref{tab:pomeron_models_table} indicates different choices of pomeron flux $f_{\mathbb{P}/p}(\xi, t)$  and pomeron set $f_{i/\mathbb{P}}(\beta, Q^{2})$, using \textsc{PYTHIA8} nomenclature, coming mostly from data-driven measurements. For the photon-induced case on the other hand, two alternative photon flux parameterizations implemented in \textsc{PYTHIA8} were considered: the Budnev model~\cite{BUDNEV} and a virtuality-dependent flux based on the Drees–Zeppenfeld approximation~\cite{Drees-Zeppenfeld}, which employs a dipole form factor to suppress contributions from high virtualities. Both of these alternative models are implemented at LO accuracy while default mode operates at NLO.\\

\begin{table}[h!]
\centering
\renewcommand{\arraystretch}{1.2}
\resizebox{1.0\columnwidth}{!}{%

\begin{tabular}{@{}lcc@{}}
\toprule
Data Driven Fit \quad \quad & Pomeron Flux \quad \quad  & Pomeron Set \\ \midrule
Default  & 1 & 6 \\
H1-A-NLO & 6 & 3 \\
H1-B-NLO & 7 & 6 \\
H1-B-LO & 7 & 4 \\
H1-AB-LO & 8 & 6 \\
\bottomrule
\end{tabular}
}
\caption{\label{fig:schemes} \justifying 
Summary of the pomeron schemes implemented. In \textsc{PYTHIA8} the parton distribution of pomeron choice corresponds to the \emph{pomeron set} parameter, used to simulate single-diffractive \( t\bar{t} \) production.
}
\label{tab:pomeron_models_table}
\end{table}


\noindent For $p'\mathbb{P}p\rightarrow p't\bar{t}X$ production, the total cross section receives contributions from both \( q\bar{q} \to t\bar{t} \) and \( gg \to t\bar{t} \) subprocesses, with the $gg$ channel dominating by approximately an order of magnitude. In contrast, the photon-induced process is driven primarily by $\gamma g$ fusion, with a smaller contribution from $\gamma q$ interactions. The corresponding predictions for both production modes, evaluated at a center-of-mass energy of 13~TeV and for different pomeron schemes, are summarized in Table~\ref{tab:ttbar_cross_sections}. The dependence of the total cross section on the pomeron parametrization reflects the sensitivity of the result to the chosen scheme. For single-diffractive top-quark pair production, the prediction obtained with the default pomeron scheme is given by Eq.~\ref{eq:sigma_Pp}.  

\begin{equation}
    \sigma_{p'\mathbb{P}p} = 12.03 \pm 0.16~\text{(stat)}{+7.52}~\text{(syst)}~\text{pb}
    \label{eq:sigma_Pp}
\end{equation}

\noindent In addition to the statistical uncertainty, diffractive predictions involve large \emph{pomeron-model} uncertainties stemming from the phenomenological treatment of DPDFs. Because hard diffractive $pp$ scattering lacks a rigorous factorization theorem, these effects were treated as scheme-dependent systematics obtained from a scan over different pomeron parameterizations. The systematic component of the uncertainty corresponds to the maximum variation of $\sigma_{p'\mathbb{P}p}$ observed across all pomeron schemes. This variation was consistently positive with respect to the default prediction and represents a scheme-dependent systematic uncertainty that clearly dominates over the statistical precision, indicating that the predicted rate for pomeron-induced case is mainly determined by the assumptions on the pomeron flux and the DPDF parametrization. The largest deviation of $\sigma_{p'\mathbb{P}p}$ relative to the default configuration occurs in the H1-AB-LO pomeron scheme, which employs a user-defined Regge flux implemented as the default flux setup in the simulation while keeping the partonic content of the pomeron identical to the default scheme. In this configuration, the predicted cross section increases by approximately $62.5\%$ compared to the default value. This behavior demonstrates that variations in the \textit{pomeron flux} have a significantly stronger impact on the normalization of $\sigma_{p'\mathbb{P}p}$ than changes in the \textit{pomeron set}, when the partonic structure of the pomeron is held fixed.\\


\noindent For photon-induced top-quark pair production, the total cross section obtained with \textsc{MadGraph5\_aMC@NLO} is 

\[
\sigma_{p'\gamma p} = 1.395 \pm 0.006~\text{(stat)}{-0.129}~\text{(syst)}~\text{pb}.
\]

\noindent The systematic uncertainty was estimated from the largest deviation with respect to the \textsc{PYTHIA8} Budnev and Drees--Zeppenfeld flux predictions, which consistently yielded lower cross sections. The maximum downward deviation was 9.2\%, indicating a flux-model dependence smaller than in the pomeron case by approximately one order of magnitude.\\

\begin{table}[h!]
\centering
\resizebox{1.0\columnwidth}{!}{%

\begin{tabular}{lccc}
\hline
\textbf{Pomeron Scheme} & $q\bar{q} \rightarrow  t\bar{t}$ [pb] & $gg \rightarrow t\bar{t}$ [pb] & Total $\sigma$ [pb] \\
\hline
Pythia8 Default        & $1.38 \pm 0.04$ & $10.65 \pm 0.11$ & $12.03 \pm 0.12$ \\
H1-A-NLO       & $1.94 \pm 0.06$ & $15.91 \pm 0.17$ & $17.85 \pm 0.18$ \\
H1-B-NLO       & $1.96 \pm 0.05$  & $13.49 \pm 0.14$ & $15.45 \pm 0.15$ \\
H1-B-LO        & $1.78 \pm 0.05$ & $11.85 \pm 0.13$ & $13.63 \pm 0.14$ \\
H1-AB-LO       & $2.41 \pm 0.07$ & $17.14 \pm 0.18$ & $19.55 \pm 0.19$ \\
\hline
\end{tabular}
}

\vspace{1em}

\resizebox{1.0\columnwidth}{!}{%

\begin{tabular}{lc}
\hline
\textbf{Photon Scheme} \quad \quad \quad \quad \quad \quad \quad \quad  \quad \quad \quad \quad &  Total $\sigma$ [pb] \\
\hline
\textsc{MadGraph} W. W. NLO & $1.395 \pm 0.006$ \\
\textsc{PYTHIA8} Budnev LO             & $1.266 \pm 0.001$ \\
\textsc{PYTHIA8} Drees–Zeppenfeld LO    & $1.380 \pm 0.004$ \\
\hline
\end{tabular}
}
\caption{\label{tab:ttbar_cross_sections}
\justifying
Total cross sections for \( t\bar{t} \) production at \( \sqrt{s} = 13~\text{TeV} \) from the pomeron and photon schemes. 
The uncertainties correspond to statistical uncertainties.
}
\end{table}

\begin{figure}[h!]
\centering
\includegraphics[width=\columnwidth]{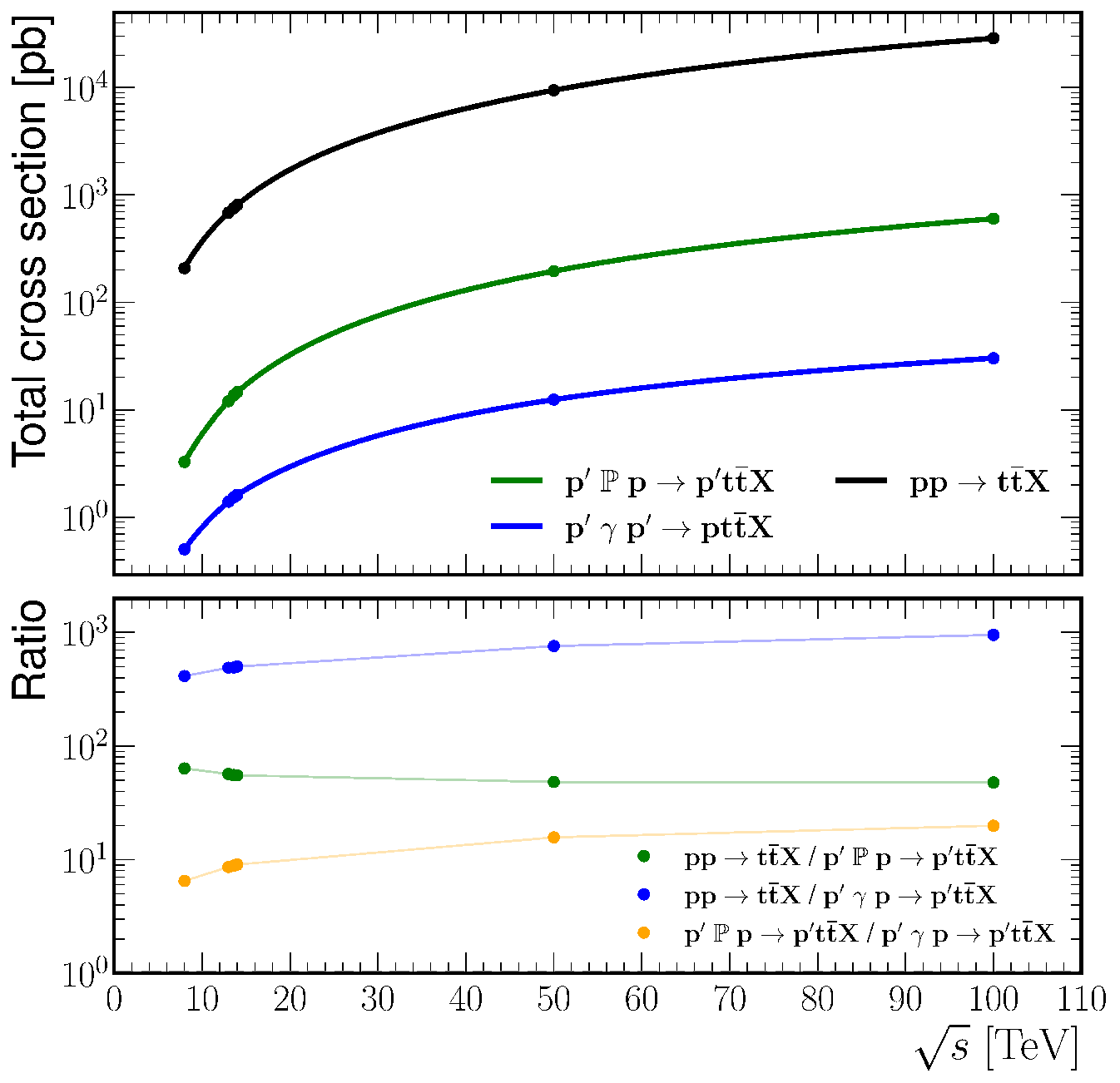}
\caption{\label{fig:xsec_vs_cem}
\justifying
Total cross sections for $t\bar{t}$ production as a function of $\sqrt{s}$, comparing non-peripheral $t\bar{t}X$ production (black) with semi-exclusive mechanisms via pomeron exchange (green) and photon exchange (blue). The lower panel shows the corresponding ratios, highlighting the stronger energy dependence of the diffractive contribution at high $pp$ center-of-mass energies.
}
\end{figure}

\noindent Differences arise from their treatment of photon virtuality and proton elastic form factors, rather than from partonic assumptions. Hence, photon-induced top-quark pair production constitutes a cleaner and more stable reference process, serving as a contrast to the scheme-dependent diffractive prediction. As mentioned above, the contributions to \( t\bar{t} \) production in $pp$ collisions with two dissociating protons that arises from \( gg \) fusion and \( q\bar{q} \) initial state interactions. These are the main backgrounds to both the pomeron and photon induced production processes. Using the default configurations indicated above for pomeron and photon induced \( t\bar{t} \) semi-exclusive production, a comparison  is performed against non-peripheral \( t\bar{t}X \) events between \( \sqrt{s} = 13~\mathrm{TeV} \) and \( \sqrt{s} = 100~\mathrm{TeV} \), which are the energy regimes at current LHC and Future Circular Collider (FCC), eras respectively. As can be seen in Figure~\ref{fig:xsec_vs_cem}. Production rates increase by a factor of $\sigma_{p'\mathbb{P}p}(100~\mathrm{TeV}) / \sigma_{p'\mathbb{P}p}(13~\mathrm{TeV}) \approx 50$ and $\sigma_{p'\gamma p}(100~\mathrm{TeV})  / \sigma_{p'\gamma p}(13~\mathrm{TeV}) \approx 21.6$, for pomeron- and photon-induced processes, respectively, after increasing the $pp$ collision energy from LHC to FCC regimes.\\

\noindent At 13\,TeV the contributions from the default pomeron and photon schemes were found to be at the level of 1.8\% and 0.2\%, respectively, relative to non-peripheral $t\bar{t}X$ events production rate. The relative fraction of the pomeron increases to $2.1\%$  and decreases to 0.01\% for the photon-induced case. The ratio between both semi-exclusive production modes rates (pomeron-photon) increases from $8.6$ at $13\,\text{TeV}$ to nearly $20$ at $100\,\text{TeV}$, indicating a steeper energy scaling in the diffractive case. Apart from the general rate increase for all \( t\bar{t} \) production modes between energy regimes, this trend suggests a particular relative increase in the diffractive production mode with respect to photo-induced production relevant for FCC projections. The faster rise might be linked to the rapid growth of pomeron gluon densities with energy with respect to the photon flux increase, reflecting in the difference in scaling behavior between the two semi-exclusive processes $\mathbb{P}p$ and $\gamma p$. Such contrasting behaviors reflect the different QCD and QED dynamics involved in each mechanism and suggest that treating them separately may be relevant for projections at future collider energies.\\


\section{\label{sec:level3} Pomeron/photon energy fractions and parton distributions}

In accordance with the IS formalism introduced in Section 2, the diffractive kinematics were interpreted in terms of the pomeron energy fraction \(\xi \equiv x_{\mathbb{P}}\) carried by the color singlet exchange. In calculations, \(x_{\mathbb{P}}\) was computed as:

\vspace{0.2cm}

\begin{equation}
    x_{\mathbb{P}} = \frac{E_{\mathbb{P}}}{E_{\text{beam}}},
\end{equation}

\begin{figure}[h!]
\centering
\includegraphics[width=\columnwidth]{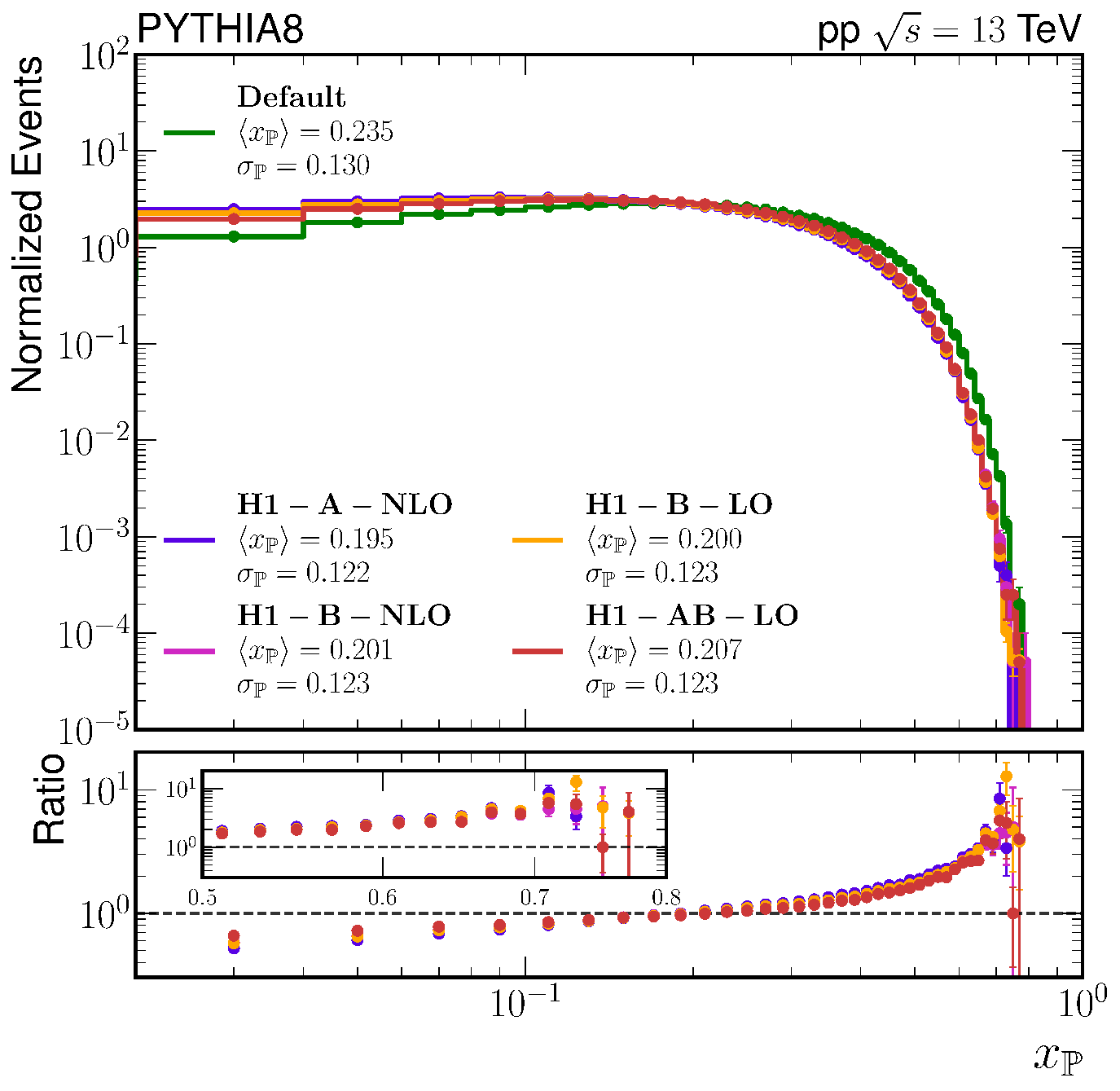}
\caption{\label{fig:ef_pomeron_models} \justifying
Normalized energy fractions distributions $x_{\mathbb{P}}$ for the five pomeron schemes defined in Table~\ref{tab:pomeron_models_table}, compared to the default model. The lower panel shows the ratio of the default scheme to each systematic variation.}
\end{figure}

\vspace{0.2cm}

\noindent where \(E_{\text{beam}} = 6500~\text{GeV}\) and \(E_{\mathbb{P}}\) denoted the pomeron energy. Accordingly, \(x_{\mathbb{P}}\) quantified the fractional proton energy transferred through the diffractive exchange, consistent with the variable $\xi$ used in the DPDF and cross-section expressions above. Figure~\ref{fig:ef_pomeron_models} shows the normalized energy fraction distributions for the five different pomeron schemes. The pomeron energy fraction with respect to the initial proton as seen in the figure extends within $0 < x_{\mathbb{P}} < 0.79$ domain. These distributions exhibit non-negligible pomeron scheme dependence, both in their shape and in the kinematic partitioning across the low ($x_{\mathbb{P}} < 0.1$), mid ($0.1 < x_{\mathbb{P}} < 0.5$), and high ($x_{\mathbb{P}} > 0.5$) energy domains. The correspondence of the different pomeron schemes with the default was evaluated through the Kolmogorov--Smirnov (KS) test implemented with \textsc{SciPy}~\cite{SciPy2020}.\\

\noindent As shown in Table~\ref{tab:xp_ks_fractions}, the most pronounced shape difference relative to the default pomeron scheme was observed in the H1-A-NLO configuration, which yielded the highest KS value (0.1266). This indicates a significant deviation from the default shape, with a visible shift of the $x_{\mathbb{P}}$ spectrum toward lower values, reducing the average $\langle x_{\mathbb{P}} \rangle$ from 0.235 to 0.195. The corresponding standard deviation values, however, remain comparable across all schemes. Comparable performance was found for H1-B-LO, H1-B-NLO, and H1-AB-LO, with the latter giving the lowest KS value (0.0858) and thus the closest agreement with the default distribution. These shape differences are reflected in the event fractions across $x_{\mathbb{P}}$ regions: H1-A-NLO produced the largest enhancement at low $x_{\mathbb{P}}$, while all schemes retained the dominant yield in the mid-$x_{\mathbb{P}}$ range. The default scheme showed the highest contribution in the high-$x_{\mathbb{P}}$ domain, with only 3.1\% of the events.
 
\begin{table}[ht]
\centering
\resizebox{1.0\columnwidth}{!}{%

\renewcommand{\arraystretch}{1.2}
\begin{tabular*}{\linewidth}{@{\extracolsep{\fill}}lcccc}
\toprule
\makecell[l]{\textbf{Pomeron} \\ \textbf{Scheme}} & 
\textbf{KS} & 
\makecell{\textbf{Low}  (\%)} & 
\makecell{\textbf{Mid}  (\%)} & 
\makecell{\textbf{High} (\%)} \\
\midrule
Default    & ---            & 16.44 & \textbf{80.46} & \textbf{3.10} \\
H1-A-NLO   & \textbf{0.1266} & \textbf{26.22} & 72.35 & 1.43 \\
H1-B-NLO   & 0.1046         & 24.33 & 74.12 & 1.55 \\
H1-B-LO    & 0.1083         & 24.49 & 74.00 & 1.52 \\
H1-AB-LO   & 0.0858         & 22.30 & 76.10 & 1.60 \\
\bottomrule
\end{tabular*}
}
\caption{\label{tab:xp_ks_fractions} \justifying
Kolmogorov--Smirnov (KS) values with respect to the default pomeron scheme, and event fractions in low, mid, and high-$x_{\mathbb{P}}$ regions.}
\end{table}

\begin{figure}[h!]
\centering
\includegraphics[width=\columnwidth]{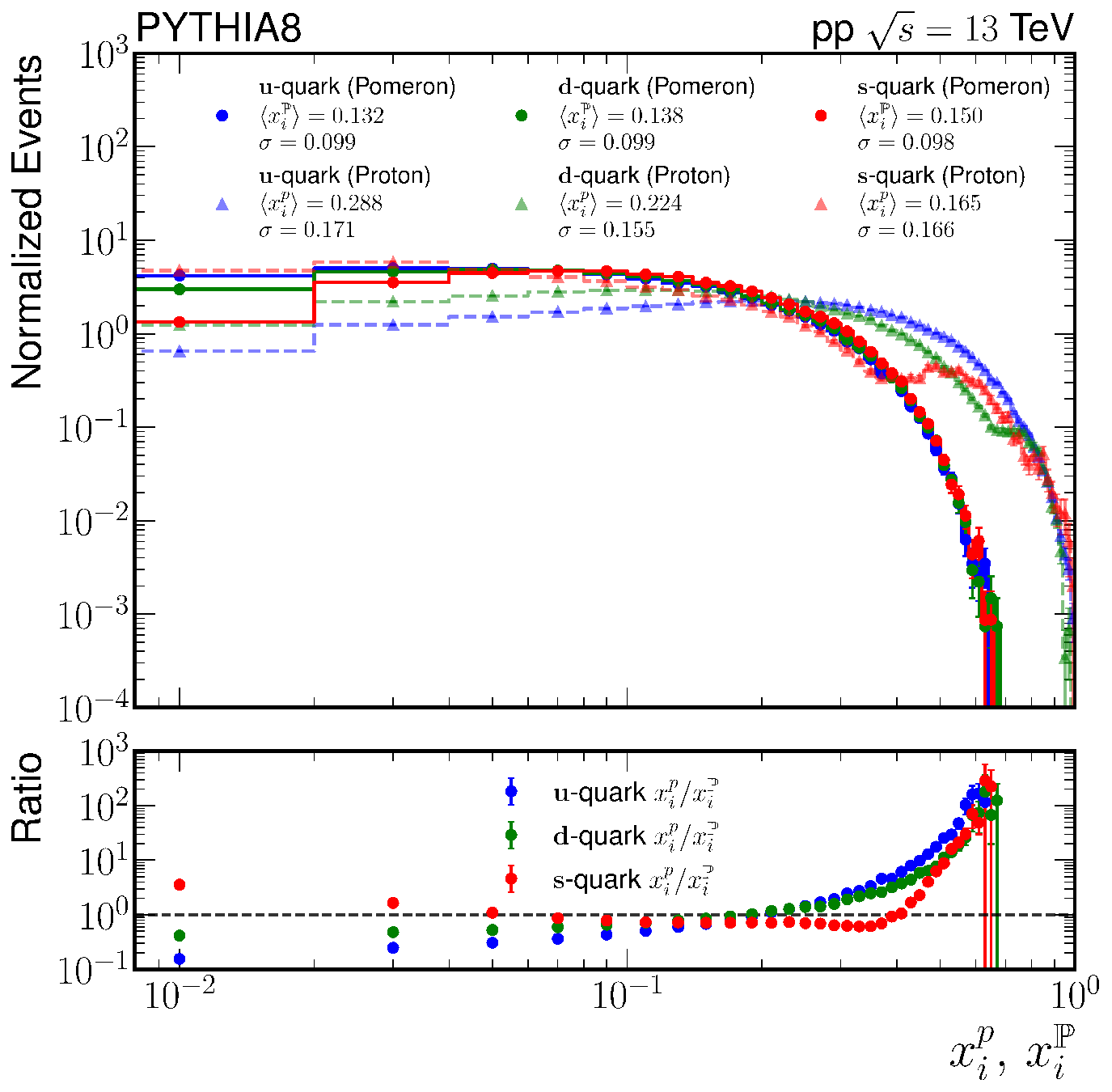}
\caption{\label{fig:quark_flavor_ratios_proton_pomeron} \justifying 
Normalized energy fraction distributions ($x_{i}^{p}$ and $x_{i}^{\mathbb{P}}$) for $u$, $d$, and $s$ quarks originating from the intact proton and from the pomeron, based on default pomeron scheme.}
\end{figure}

\begin{figure}[h!]
\centering
\includegraphics[width=\columnwidth]{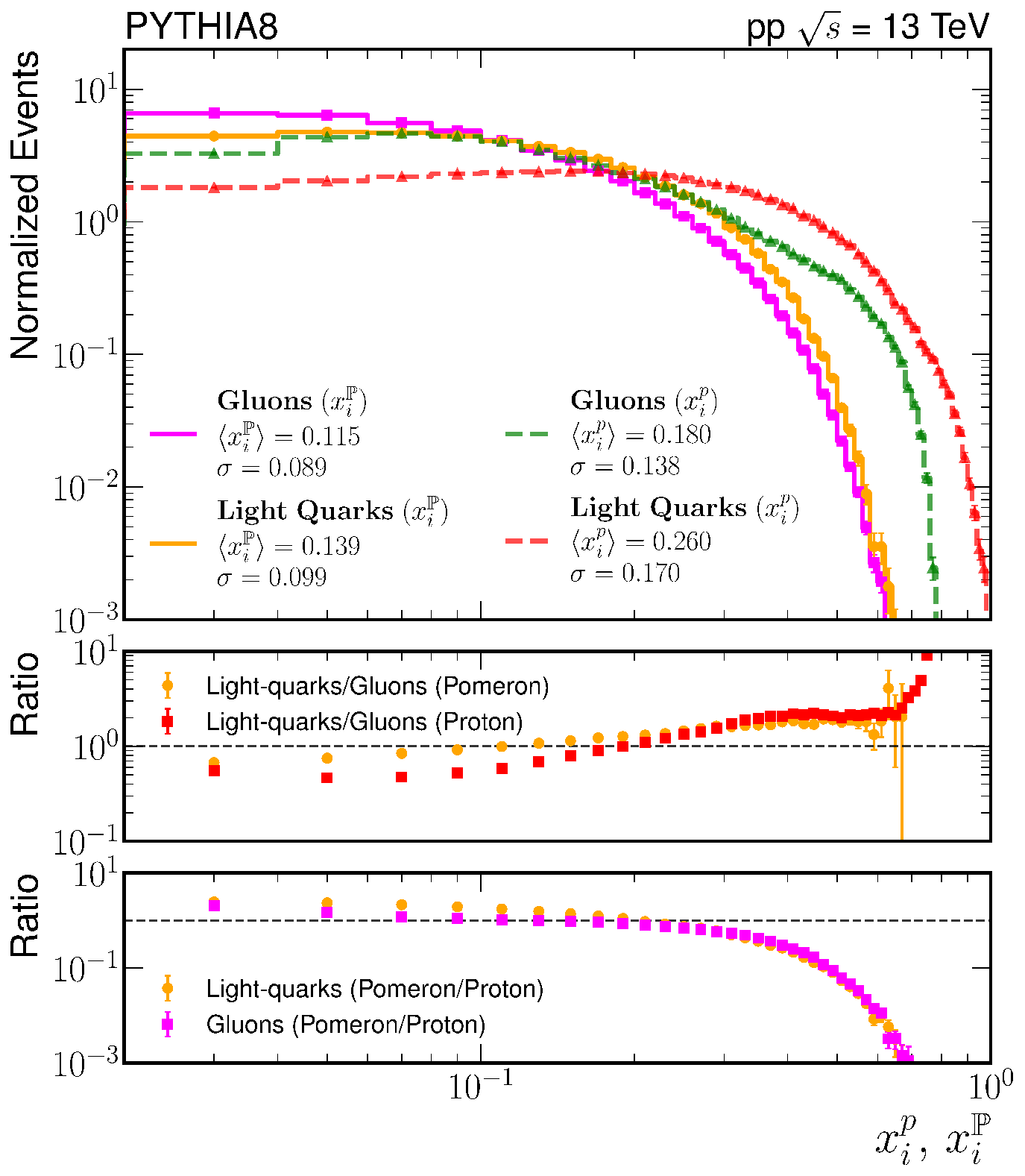}
\caption{\label{fig:energy_fraction_pom_prot} 
\justifying 
Normalized energy fraction distributions for partons originating from the pomeron and proton under the default pomeron scheme. 
Results are shown for light quarks (\(i = u+d+s\)) and gluons (\(i = \text{gluon}\)).}
\label{fig:ef_photon_quarks_gluons_pomeron_proton}
\end{figure}

\noindent In the parton model, the internal structure of hadrons such as the proton are described in terms of constituent quarks and gluons, each carrying an energy fraction $x$ of the whole partonic structure. A similar approach is followed in the non-perturbative approach used for the pomeron in this analysis. The diagram at the top of Figure~\ref{fig:diagrams_photon_pomeron} indicates a breaking proton side and a pomeron side both acting as sources of initial state partons.  In this context, the energy fraction carried by a parton $i$ from the breaking proton is defined as $x_i^{p} = \frac{E_i^{p}}{E_{\text{beam}}}$, where $E_i^{p}$ is the energy of the parton $i$ (with $i = u, d, s$). For the pomeron side, the energy fraction of the parton \( i \) associated with the pomeron is defined as \( x_i^{\mathbb{P}} = E_i^{\mathbb{P}} / E_{\text{beam}} \), which, in terms of the diffractive variables introduced above, satisfies \( x_i^{\mathbb{P}} = \beta\,\xi \). The normalized $x_i^{p}$, $x^{\mathbb{P}}_i$ distributions of $u$, $d$, and $s$ quark flavors and gluons using the pomeron default scheme are shown in Figure~\ref{fig:quark_flavor_ratios_proton_pomeron} and~\ref{fig:ef_photon_quarks_gluons_pomeron_proton}. First visible contrasting difference between $x_i^{p}$ and $x^{\mathbb{P}}_i$ seen in Figure~\ref{fig:quark_flavor_ratios_proton_pomeron} is that the latter is limited to the pomeron maximum relative energy to the proton shown in Figure~\ref{fig:ef_pomeron_models} of about 0.79, while $x_i^{p}$ spectrum spans the entire range, with the $u$ quark component dominating at high $x$.\\

\noindent Table~\ref{tab:parton_fractions} summarizes key features of the flavor and partonic composition of the pomeron and the proton. The total flavor fractions indicate that the proton is dominated by $u$-quarks ($\sim62\%$), with a lower $d$-quark fraction and a small contribution from $s$-quarks ($\sim6\%$). In contrast, the pomeron shows a more balanced scenario among $u$, $d$, and $s$ quark contributions, each contributing roughly $30$--$37\%$, this might be associated with the valence quark structure present within the proton with dominant $u$-quark flavor. The gluon saturation level is stronger within the pomeron with a contribution of $\sim80\%$, while the proton exhibits a more balanced structure with contributions of $55\%$ for gluons and $45\%$ for light quarks. Though the gluon dominance increases to $\sim98\%$ for the pomeron at FCC energy regime.\\

\noindent The pomeron-to-proton ratio of normalized entries in $x_i^{p}$ and $x^{\mathbb{P}}_i$ distributions for individual quark flavors across low, mid and high range regions reveal that at low $x$, the $u$ and $d$ quarks are enhanced in the pomeron (ratios $>1$), whereas $s$-quarks are closer to unity. At high $x$, all flavor ratios are strongly suppressed in the pomeron with respect to proton. In the mid-$x$ region, the $s$-quark ratio stands out as the largest among the three. Another aspect to consider is the relative contributions between light quarks to gluons in the different range regions. The light-quark to gluon ratio is below one at low $x$ for both sources, but increases with $x$, exceeding unity in the pomeron at mid and high $x$. Finally, the pomeron-to-proton ratios for light quarks and gluons in final row at the table follow a similar pattern: they are above one at low $x$, drop below one at mid $x$, and become nearly zero at high $x$. From Figure~\ref{fig:ef_pomeron_models} encodes the normalized shapes of the pomeron flux \( f_{\mathbb{P}/p}(x_{\mathbb{P}}) \), while Figure~\ref{fig:quark_flavor_ratios_proton_pomeron} presented the parton distributions inside the pomeron \( f_{i/\mathbb{P}}(\beta) \) together with the proton PDFs for comparison.\\\\

\begin{table}[ht]
\centering
\small
\renewcommand{\arraystretch}{1.25} 
\resizebox{\columnwidth}{!}{%
\begin{tabular}{@{}c|lccc@{}}
\toprule
\textbf{} & \textbf{Source} & $u$-quark & $d$-quark & $s$-quark \\
\midrule
\makecell[c]{\textbf{Total} \\ \textbf{flavor} \\ \textbf{fraction}} 
& Pomeron & $36.58\%^{+0.21\%}_{-0.21\%}$ & $34.25\%^{+0.21\%}_{-0.21\%}$ & $29.17\%^{+0.20\%}_{-0.20\%}$ \\
& Proton  & $61.62\%^{+0.14\%}_{-0.14\%}$ & $32.81\%^{+0.14\%}_{-0.14\%}$ & $5.57\%^{+0.07\%}_{-0.07\%}$ \\
\midrule
\textbf{} & \textbf{Source} & Light-quarks & Gluons & \\
\midrule
\makecell[c]{\textbf{Partonic} \\ \textbf{content}} 
& Pomeron & $19.72\%^{+0.08\%}_{-0.07\%}$ & $80.28\%^{+0.07\%}_{-0.07\%}$ & \\
& Proton  & $45.19\%^{+0.10\%}_{-0.10\%}$ & $54.81\%^{+0.10\%}_{-0.10\%}$ & \\
\midrule
\textbf{} & \textbf{Flavor} & Low $x$ & Mid $x$ & High $x$ \\
\midrule
\makecell[c]{\textbf{Pom./Prot.} \\ \textbf{Ratio}} 
& $ \quad u$ & $3.314 \pm 0.025$ & $0.729 \pm 0.004$ & $0.015 \pm 0.001$ \\
& $ \quad d$ & $1.844 \pm 0.015$ & $0.798 \pm 0.005$ & $0.036 \pm 0.003$ \\
& $ \quad s$ & $0.808 \pm 0.009$ & $1.333 \pm 0.014$ & $0.032 \pm 0.003$ \\
\midrule
\textbf{} & \textbf{Region} & Pomeron & Proton & \\
\midrule
\makecell[c]{\textbf{Light-quarks/} \\ \textbf{Gluons Ratio}} 
& Low $x$       & $0.788 \pm 0.003$ & $0.534 \pm 0.002$ & \\
& Mid $x$       & $1.248 \pm 0.004$ & $1.174 \pm 0.003$ & \\
& High $x$       & $1.814 \pm 0.109$ & $2.463 \pm 0.020$ & \\
\midrule
\textbf{} & \textbf{Region} & Light-quarks & Gluons & \\
\midrule
\makecell[c]{\textbf{Pom./Prot.} \\ \textbf{LQ,G Ratio}} 
& Low $x$       & $2.259 \pm 0.011$ & $1.530 \pm 0.004$ & \\
& Mid $x$       & $0.804 \pm 0.003$ & $0.756 \pm 0.002$ & \\
& High $x$      & $0.020 \pm 0.001$ & $0.028 \pm 0.001$ & \\
\bottomrule
\end{tabular}%
}
\caption{\label{tab:parton_fractions}
\justifying Summary of quark and gluon fractions and their ratios from normalized energy fraction distributions. Includes flavor content, partonic composition, pomeron-to-proton and light-quark-to-gluon ratios in different energy fraction regions. Uncertainties correspond to 95\% confidence level and at  $\sqrt{s} = 13 \text{ TeV}$.}
\end{table}

\begin{figure}[h!]
\centering
\includegraphics[width=\columnwidth]{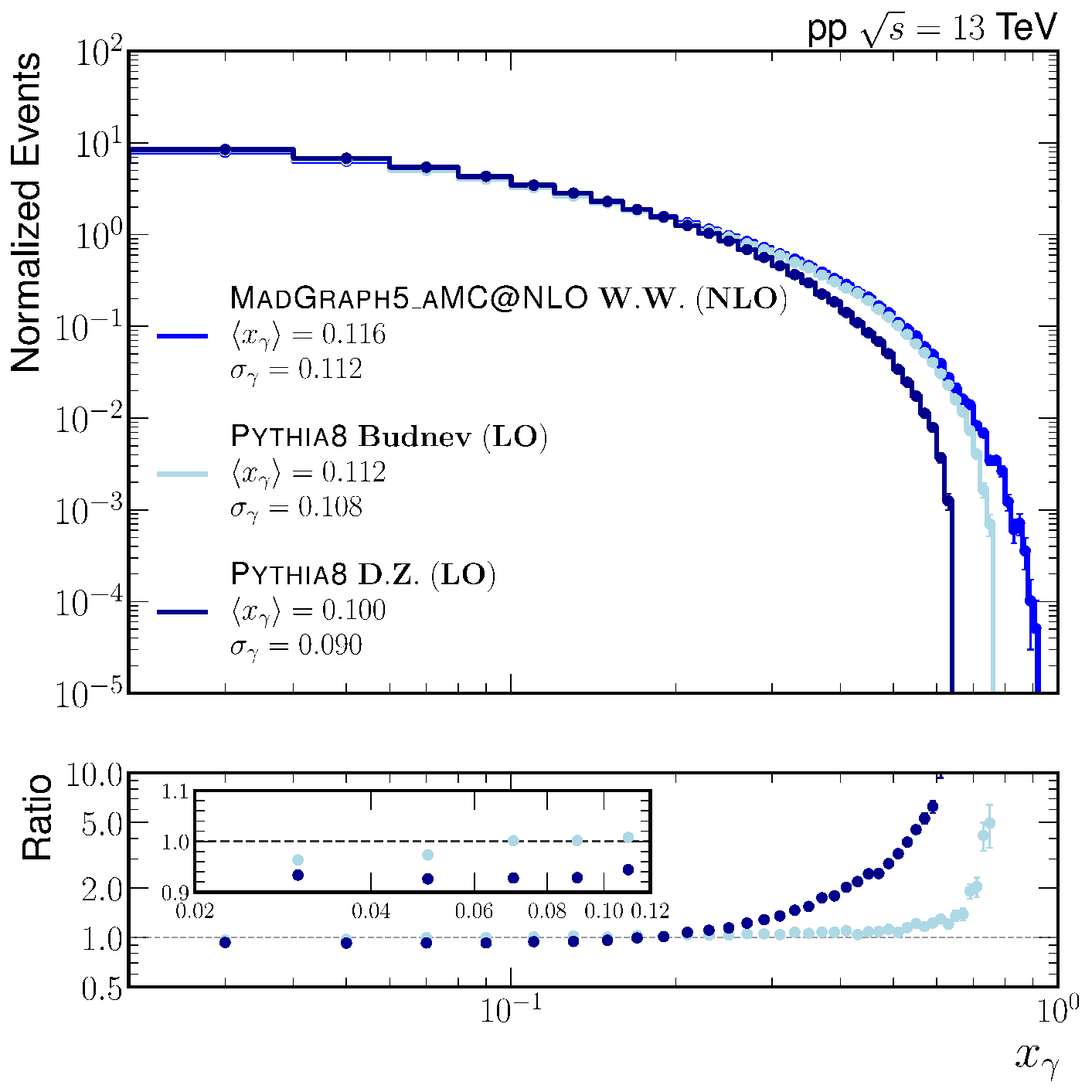} \\
\caption{\label{fig:energy_fraction_photon} \justifying Normalized energy fraction distributions for photons in W. W., Budnev and Drees–Zeppenfeld flux schemes.}
\label{fig:photonflux_epa} 
\end{figure}

\noindent In the case of photon-induced $t\bar{t}$ events, illustrated at the bottom of Figure~\ref{fig:diagrams_photon_pomeron}, Table~\ref{tab:xgamma_fractions} summarizes the event fractions across three energy regions. The corresponding normalized photon energy fraction with respect to the proton, $x_\gamma$, is presented in Figure~\ref{fig:photonflux_epa}. Three different photon flux schemes are shown, the default model specified in Section 2 and two additional scheme variations from PYTHIA8 generator. The photon flux exhibited notable differences across the three models, particularly in the low- and high-$x_\gamma$ regions. For the default EPA Weizsäcker–Williams scheme, 58.9\% of photons were found below $x_\gamma < 0.1$ and 1.1\% above $x_\gamma > 0.5$. The Budnev flux led to a softer distribution, with 60.3\% of events at $x_\gamma < 0.1$ and 0.87\% at high $x_\gamma$. The Drees–Zeppenfeld parametrisation exhibited the strongest suppression in the high-$x_\gamma$ region, with 63.0\% of events at $x_\gamma < 0.1$ and only 0.2\% at $x_\gamma > 0.5$. Overall, the photon contribution at $x_\gamma > 0.5$ remained below 1\%, while the dominant yield was concentrated at $x_\gamma < 0.1$, where deviations from the default prediction reached up to $\sim$7\%. \\

\begin{table}[ht]
\centering
\resizebox{1.0\columnwidth}{!}{%

\begin{tabular}{lccc}
\toprule
\textbf{Photon Flux} & \textbf{Low($\%$)} & \textbf{Mid}($\%$) & \textbf{High}($\%$)\\
\midrule
Weizsäcker–Williams  & 58.87 & 40.06 & 1.08 \\
Budnev               & 60.30 & 38.83 & 0.87 \\
Drees–Zeppenfeld     & 62.98 & 36.82 & 0.20 \\
\bottomrule
\end{tabular}
}
\caption{\label{tab:xgamma_fractions}
Fractions of events (\%) in the low-, mid-, and high-$x_\gamma$ regions for different photon–flux schemes.}

\end{table}

\noindent The diffractive exchange is found to be largely gluon dominated ($\sim80\%$ at 13~TeV, increasing to $\sim98\%$ at 100~TeV), while the proton side retains a pronounced high-$x$ valence-$u$ component, and $gg\to t\bar{t}$ processes dominate in diffractive production. These features, however, as indicated above, remain subject to uncontrolled modeling systematics arising from the choice of pomeron parameterization, flux normalization, and gap-survival assumptions, which can impact the relative gluon and quark contributions within the diffractive exchange. For photon-induced events, all flux models exhibit a stable low-$x_\gamma$ peak with negligible contributions above $x_\gamma>0.5$ and modest model variations (up to $\sim7\%$) concentrated at low $x_\gamma$. In single-diffractive kinematics, the variables introduced above are directly connected to experimentally measurable gap observables. The proton fractional momentum loss $\xi$ (or equivalently $x_{\mathbb P}$) determines the rapidity–gap size through $\Delta\eta_{\text{gap}} \simeq \ln(1/\xi) \simeq \ln(1/x_{\mathbb P})$, so that smaller $\xi$ values correspond to larger forward gaps and reduced particle activity~\cite{Rasmussen_2018}. The observed energy–fraction patterns provide a consistent picture of semi–exclusive top–pair production, where both the pomeron and the photon exchanges carry only a fraction of the proton energy. In the diffractive case, the exchanged pomeron is predominantly gluon–driven, and the shift of the $x_{\mathbb P}$ spectra toward lower values in some models reflects enhanced soft–gluon contributions, which enlarge the rapidity gap and suppress central activity. In photon–induced interactions, a similar behavior arises from the steeply falling $x_\gamma$ spectrum, leading to large forward gaps associated with low–$x_\gamma$ photons. Altogether, these kinematic features establish the physical basis for the gap and forward–multiplicity observables that will be discussed in Sec.~4.


\section{Rapidity gaps and particle multiplicity}

\noindent As introduced in Sec.~1, in forward physics, neutral fluxes mediated by photons or pomerons often lead to suppressed color flow, resulting in emergence of angular regions with no final state particles present in collision events as shown at the top of Figure~\ref{fig:rapidity_gap_illustration} where the left side of polar angle range is empty with no particles present in contrast with a non-peripheral $t\bar{t}X$ collision topology at the bottom. These empty regions are characterized by their pseudorapidity angular distance and are regarded as rapidity gaps.\\

\noindent The forward rapidity Gap (FRG) is defined as an empty region (no charged final state particle with transverse momentum p$_\mathrm{T}$ $>$ 200 MeV) measured with respect to the intact proton direction, which is $\eta$  equal to $-$2.4 or $+$2.4 for the case of intact proton traveling to positive or towards negative direction respectively as summarized below:\\

\vspace{-0.5cm}

\begin{align}
\mathrm{Gen.\ FRG}\ \bigl(\textnormal{positive side}\bigr) 
  &= \min_i \left| \eta_i - 2.4 \right|  \label{eq:frg_pos} \\[6pt]
\mathrm{Gen.\ FRG}\ \bigl(\textnormal{negative side}\bigr) 
  &= \min_i \left| \eta_i + 2.4 \right| . \label{eq:frg_neg}
\end{align}

\begin{figure}[h!]
    \centering
    \includegraphics[width=\columnwidth]{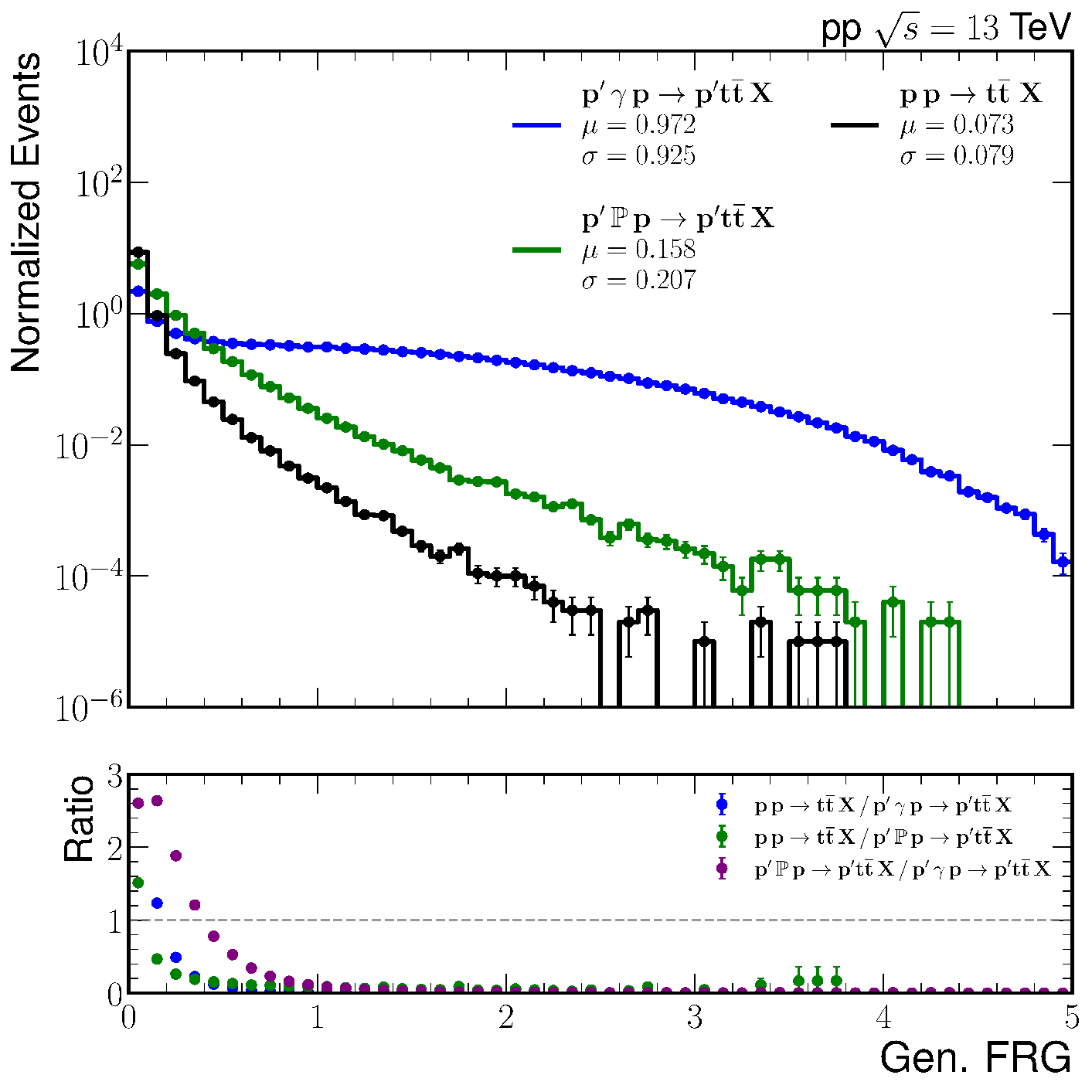}
    
    \vspace{0.5em}
    \textbf{(a)} Generator level FRG.
    
    \vspace{1.5em}
    \includegraphics[width=\columnwidth]{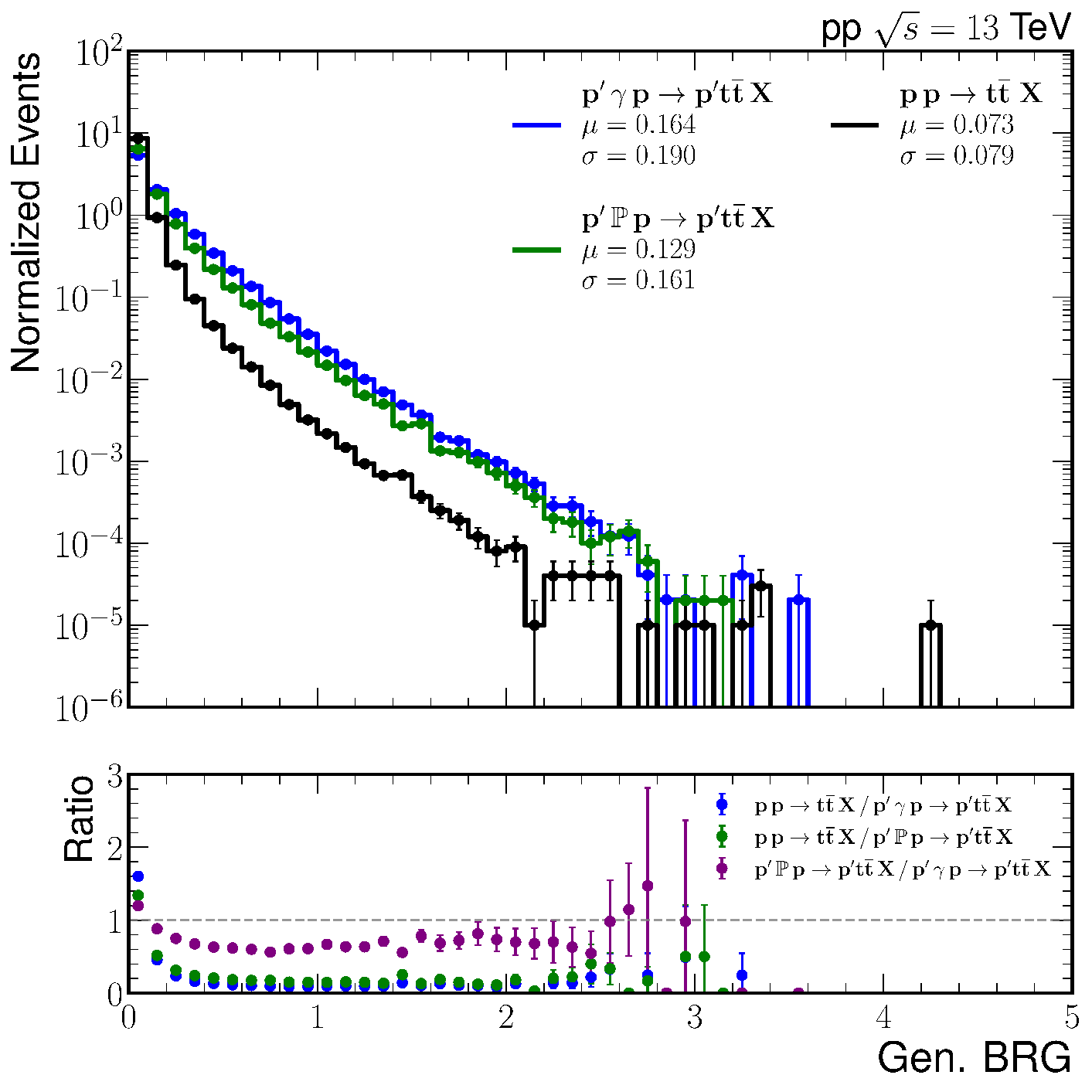}
    
    \vspace{0.5em}
    \textbf{(b)} Generator level BRG.
    
    \caption{\label{fig:dynamic_rapdity_gap_distributions} \justifying Normalized distributions of forward (a) and backward (b) rapidity gaps for photon-induced, pomeron-induced, and non-peripheral $t\bar{t}X$ selected events.}
    \label{fig:dynamic_gap}
\end{figure}

\noindent Equations~(\ref{eq:frg_pos}) and~(\ref{eq:frg_neg}) define the generator-level FRG on the positive and negative sides of the detector, respectively. The minimization runs over all final-state particles with momentum above 200\,MeV. The FRG is given by the smallest pseudorapidity distance between the detector edge e.g. CMS Tracker at $\eta = - 2.4$ and the nearest final state particle. By construction, this value corresponds to the size of the empty rapidity interval adjacent to the beam direction on each side. FRG variable is aimed to measure the activity imbalance introduced in photon or pomeron induced $t\bar{t}$ events with respect to non-peripheral $pp$ collisions~\cite{Khoze:2002nf, Tumasyan_2023}. The Gen. BRG variable was defined on the other hand with respect to breaking proton direction either $+$2.4 or $-$2.4 in $\eta$. At the top of Figure~\ref{fig:rapidity_gap_illustration}, the intact proton travels in the $+z$ direction and carries a positive large fraction of the beam momentum ($p_z > 0$), resulting in a characteristic suppression of hadronic activity in the negative pseudorapidity region. This produces an FRG, with the forward region $\eta > 0$ more populated in comparison with the $\eta < 0$ region. Figure~\ref{fig:dynamic_rapdity_gap_distributions} shows the gap distributions obtained for the three samples $p'\mathbb{P}p \rightarrow p't\bar{t}X$, $p'\gamma p \rightarrow p't\bar{t}X$ and $pp \rightarrow t\bar{t}X$. The distinction between semi-exclusive and non-peripheral $t\bar{t}X$ processes is evident in both regions in the Gen. FRG distribution. The fraction of events with a gap size $\geq 1.0$ was found to be highest in the forward region for photon-induced (41.07\%) and pomeron-induced (1.04\%) processes, compared to the non-peripheral $t\bar{t}X$ events sample (0.07\%). These indicates that the FRG is a clear signature for photon-induced events and can serve as a discrimination tool also for pomeron-induced events. For Gen. BRG (taking breaking proton as reference), the corresponding fractions above a magnitude of 1.0 for  semi-exclusive cases were smaller than Gen. FRG: 0.71\% for photon-induced, 0.47\% for pomeron-induced, and  remain the same  0.07\% for non-peripheral $t\bar{t}X$ events. These results suggest a significantly higher probability of observing a large gap in semi-exclusive processes, especially photon-induced events, in both the forward and backward regions.\\

\begin{figure}[h!]
\includegraphics[width=\columnwidth]{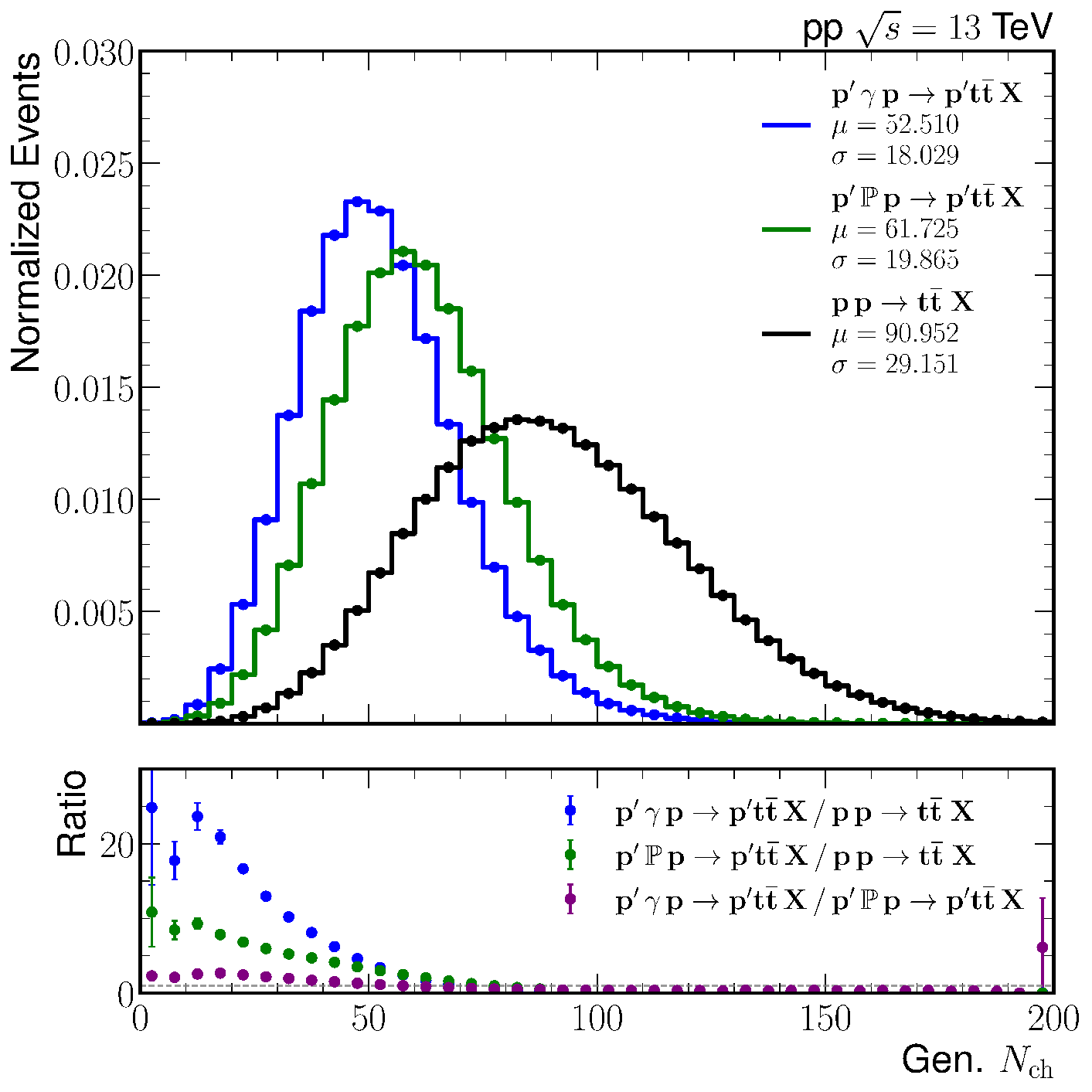}
\caption{\label{fig:nChargedTrk_comparison} \justifying Charged-particle multiplicity $N_{\text{ch}}$ at generator level for stable particles with $p_{T} > 0.4$\,GeV and $|\eta| < 2.4$.}

\end{figure}

\noindent Another aspect to characterize photon and pomeron induced modes with respect non-peripheral $t\bar{t}X$ production is to assess the amount of particles that are created in a single $pp$ collision event. For our analysis, we defined the charged particle multiplicity $N_{\mathrm{ch}}$ as number of stable charged particles  with transverse momentum $p_{T} > 0.4~\mathrm{GeV}$ and pseudorapidity within $|\eta| < 2.4$. The resulting distributions of $ \text{Generator }N_{\mathrm{ch}}$ for the three production mechanisms are shown in Fig.~\ref{fig:nChargedTrk_comparison}, allowing for a direct comparison for particle production capability in photon-pomeron induced and non-peripheral $t\bar{t}X$ events. A substantial fraction of photon-induced events (47.6\%) and pomeron-induced events (28.8\%) exhibit low charged-particle multiplicities ($N_{\text{ch}} < 50$), compared to only 6.7\% in non-peripheral $t\bar{t}X$ events. Comparatively, high-multiplicity tails ($N_{\text{ch}} > 80$) are significantly more populated in non-peripheral $t\bar{t}X$ events (62.1\%) relative to pomeron-induced (16.9\%) and photon-induced (7.1\%) events. Particle multiplicity distributions show the largest average values for non-peripheral $t\bar{t}X$ events and the lowest values for photo-induced events. In photon- and pomeron-induced scenarios, lower multiplicities are expected as a consequence of color-neutral exchange and suppressed radiation. In contrast, non-peripheral $t\bar{t}X$ events involve color exchange between the incoming partons, resulting in significantly increased particle production. This contrast in particle multiplicity distributions provides a clear distinction between the photon-induced, single-diffractive and the non-peripheral $t\bar{t}X$ production modes. These differences motivate Sec.~5, which examines how semi-exclusive and non-peripheral $t\bar{t}X$ topologies influence detector-level observables.

\section{\label{sec:level5}Detector level observables and selection efficiency} 

\begin{figure}[h!]
    \centering
    \includegraphics[width=\columnwidth]{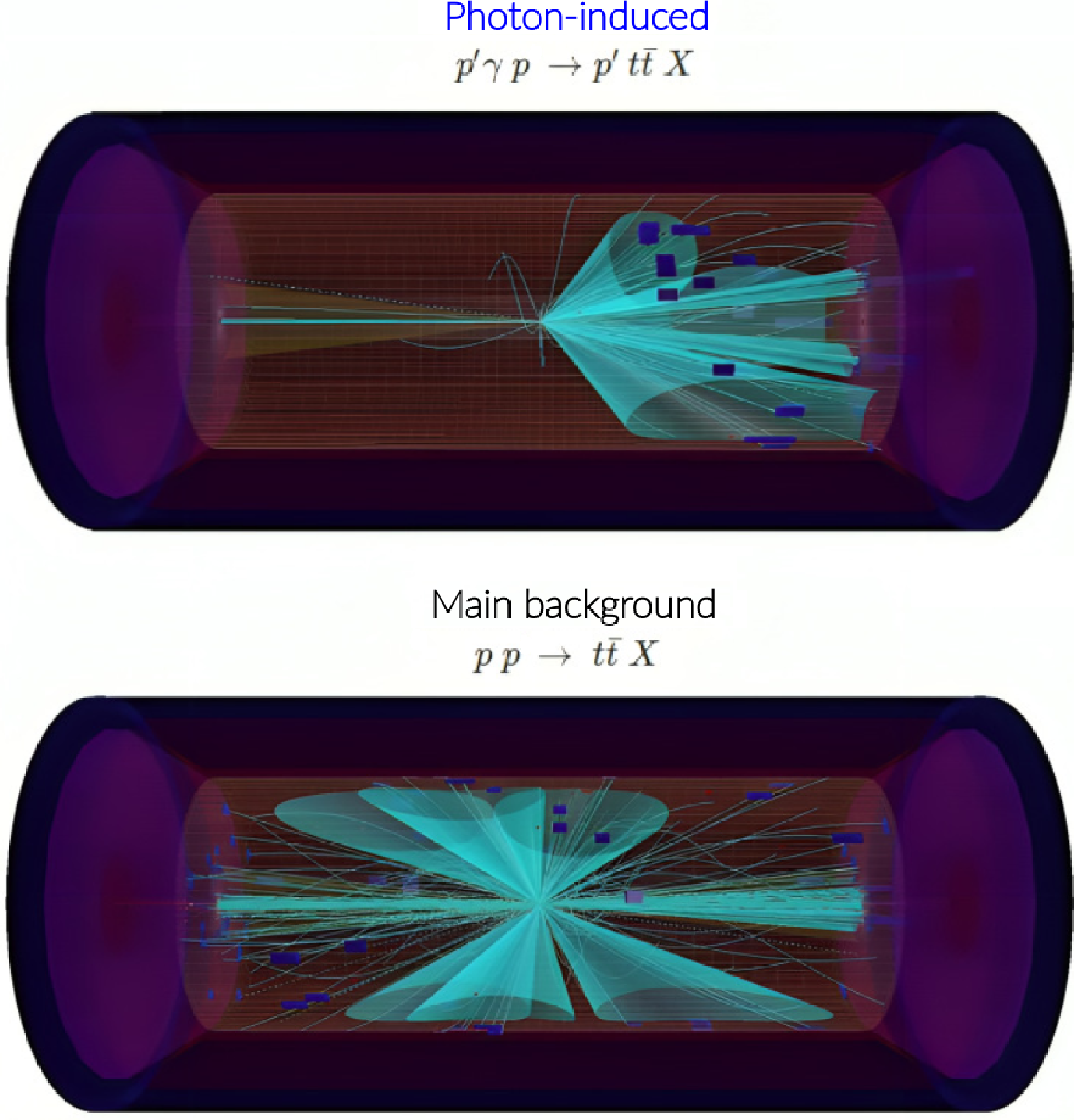}
    \vspace{0.5em}

    \caption{\label{fig:delphes_simulation} \justifying 
    Detector-level event visualization with \textsc{Delphes} using the CMS configuration. The top panel shows a photon-induced process ($p'\,\gamma\,p \rightarrow p'\,t\bar{t}\,X$) with forward activity aligned with the initial-state photon ($p_z > 0$). The bottom panel displays a non-peripheral $t\bar{t}X$ event ($p\,p \rightarrow t\bar{t}\,X$) with higher and more balanced activity throughout the detector.
    }

\end{figure}

To evaluate the impact of detector effects on the discrimination between pomeron-induced, photon-induced and standard non-peripheral $t\bar{t}X$ production modes, the generated samples were processed through a fast detector simulation with \textsc{Delphes 3}~\cite{delphes}. The CMS detector configuration was used, incorporating the most relevant detector effects, including reconstructed track efficiencies, calorimeter granularity, jet identification, $b$-tagging performance, missing transverse energy (MET), lepton identification and hadronic forward (HF) towers, among others. An illustration of the resulting detector level topologies for photon-induced and non-peripheral $t\bar{t}X$ events are shown in Fig.~\ref{fig:delphes_simulation}, revealing clear topological contrasts between semi-exclusive and non-peripheral $t\bar{t}X$ processes, which serve as the basis for selecting discriminating variables.\\

\begin{figure}[!h]
\includegraphics[width=\columnwidth]{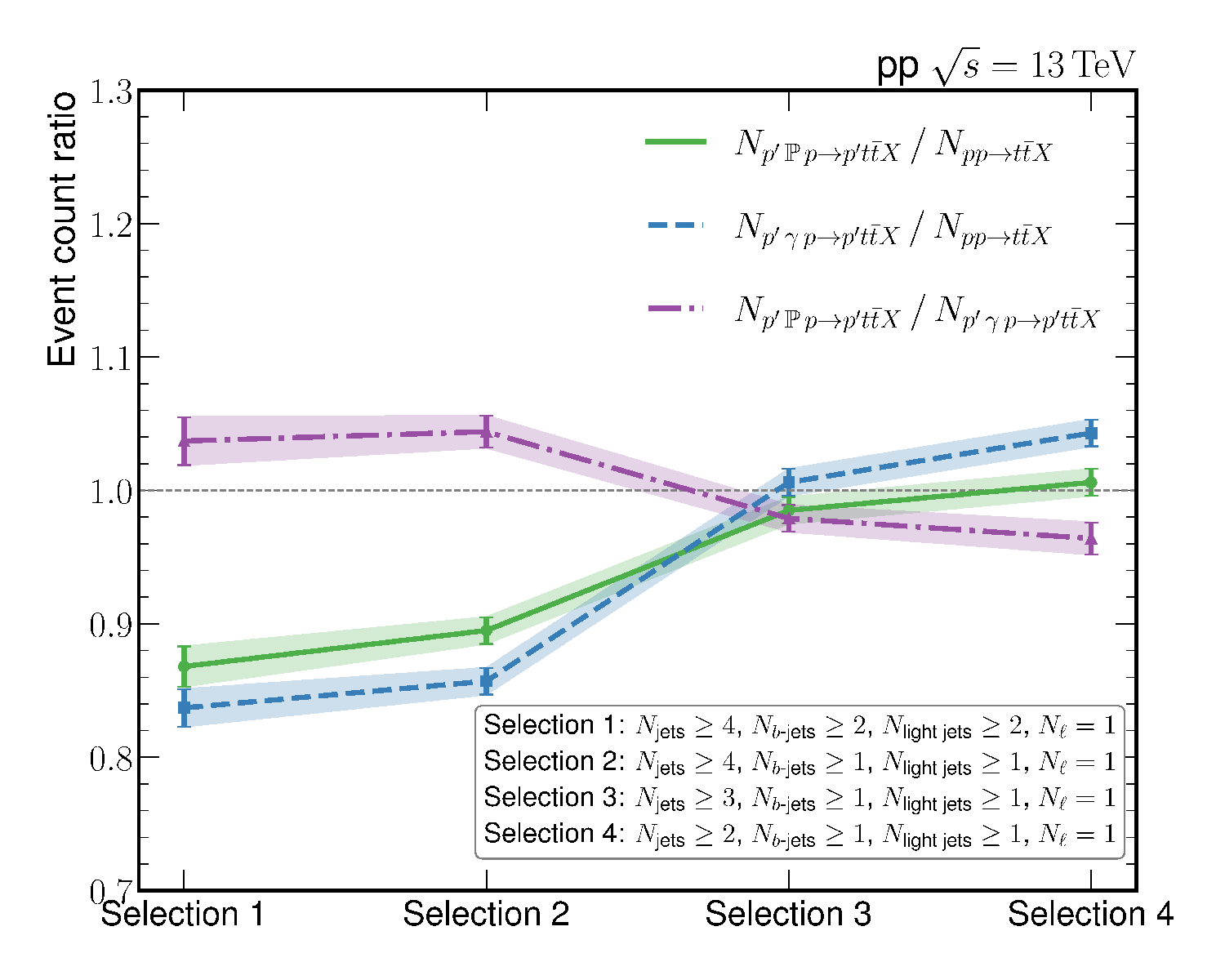}
\caption{Events ratio for different object selection.}
\label{fig:object_selection_ratio_with_band}
\end{figure}

\noindent In order to exploit topological differences, object selection criteria were first applied to identify semi-leptonic $t\bar{t}$ events across the three production modes. The resulting selection efficiency is essential to suppress non-peripheral $t\bar{t}X$ background and enhance the relative presence of photon- and pomeron-induced processes. The object selection procedure was based on jet and lepton multiplicity requirements, reflecting typical topologies arising from $t\rightarrow Wb \rightarrow bq\bar{q}'$ and $t\rightarrow bW \rightarrow bl\nu$ decays following similar procedures to~\cite{CMS:2023ebf}. Four distinct sets of criteria were imposed, varying the minimal number of total jets, $b-$tagged jets, light-flavor jets, and isolated leptons. For each selection, the ratio of events obtained after selection between production modes were computed with their respective uncertainties. These yield ratios quantify the level of significance of semi-exclusive signal with respect to the non-peripheral $t\bar{t}X$ noise and are shown in Figure~\ref{fig:object_selection_ratio_with_band}, where $N_{\text{process}}$ stands for number of events that passed each object-level selection indicated in the horizontal axis. The selection requirements are relaxed further towards the horizontal axis right direction by lowering the required minimum number of jets, $b$-jets, and light-jets, the $N_{p'\mathbb{P}p/ \text{non-peripheral }t\bar{t}X}$ and $N_{p'\gamma p/\text{non-peripheral }t\bar{t}X}$ ratios were observed to increase, reflecting a higher significance of the semi-exclusive signal. On the other hand, the $N_{p'\mathbb{P}p}/N_{p'\gamma p}$ ratio was found to decrease, indicating a relative enhancement of photon-induced events in lower-multiplicity selections. Though it is priority to keep the discrimination against non-peripheral $t\bar{t}X$ noise higher to enhance at first the discovery potential. Based on the selection efficiencies, the optimal selection was identified as $N_{\text{jets}} \geq 2$, $N_{\text{b-jets}} \geq 1$, $N_{\text{light-jets}} \geq 1$, and $N_{\text{lepton}} = 1$ corresponding to selection 4.\\

\begin{figure}[!h]
\includegraphics[width=\columnwidth]{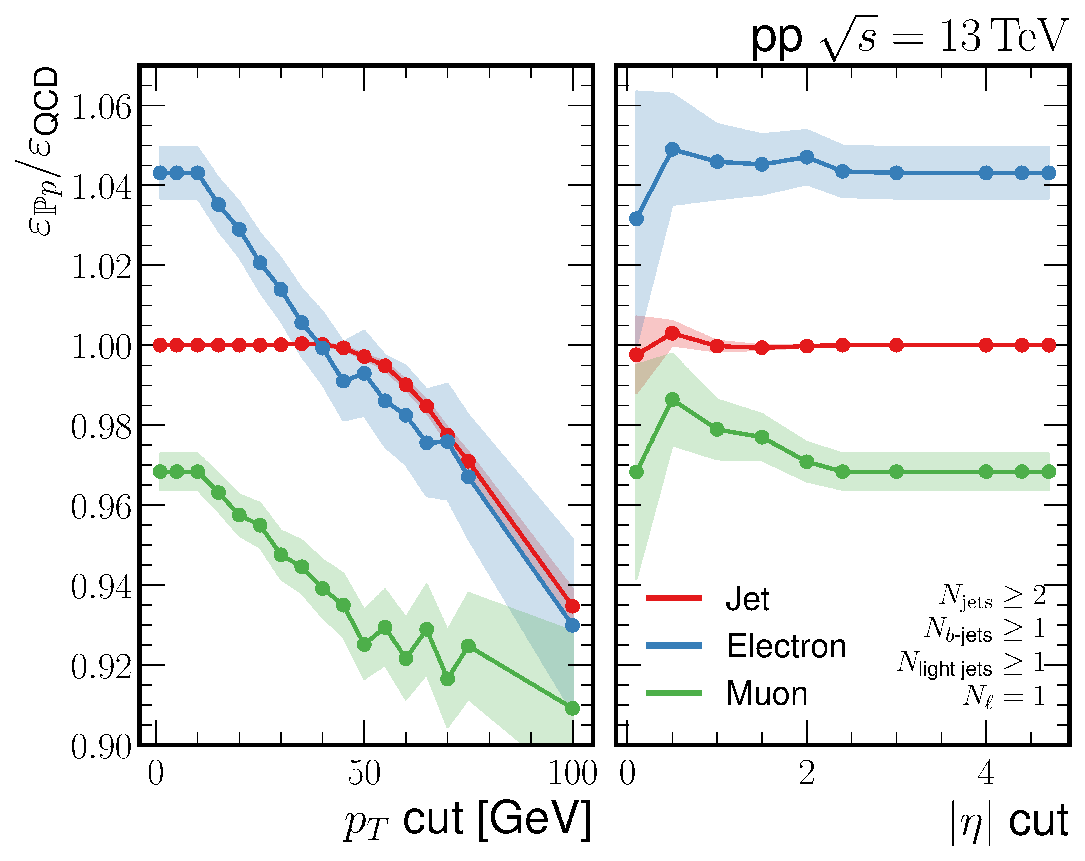}
\caption{ \justifying Efficiency ratio for different selections of jets, electron and muon objects.}
\label{fig:sb_ratio_pt_eta}
\end{figure}

\noindent The signal efficiency ($\varepsilon_{p'\mathbb{P}p}$), can be further optimized by analyzing different selection requirements for the selected objects observables such as $p_\mathrm{T}$ and $|\eta|$ as they are measured by the detector. Figure~\ref{fig:sb_ratio_pt_eta} shows the ratio $\varepsilon_{p'\mathbb{P}p} / \varepsilon_{\text{non-peripheral }t\bar{t}X}$  as a function of $p_\mathrm{T}$ and $|\eta|$ thresholds applied to jets, electrons, and muons. Thresholds were defined as $p_\mathrm{T}$ $> x$ with $x \in (0, 1000]$ GeV (left panel) and $|\eta| < x$ with $x \in [0, 4.7]$ (right panel). The vertical axis represents the relative selection efficiency with respect non-peripheral $t\bar{t}X$ background, highlighting regions where the signal efficiency is enhanced. A clear decline in the efficiency ratio is observed for electrons and muons beyond transverse momentum thresholds of approximately 10 GeV, while jets maintain a near-unity ratio up to 50 GeV. In the pseudorapidity domain, the ratio remains flat for jets across the explored $|\eta|$ range and shows only slight variations for leptons. These behaviors reflect the differing acceptance profiles and kinematic features of the underlying processes, guiding the definition of object-level thresholds adopted in the final selection. Additionally, the optimal selection for missing transverse energy was identified as $E_T^{\text{miss}} > 20$ GeV, while charged tracks were required to satisfy $p_T > 0.4$ GeV and $|\eta| < 2.4$. The full set of optimized detector-level selection thresholds are summarized in Table~\ref{tab:detector_selection}. \\

\noindent At detector level rather than using the intact proton as a reference to measure the appearance of gaps, FRG is measured with respect to $\eta$ $=$ $-$2.4 and BRG with respect $\eta$ $=$ $+$2.4, so similar plots for these two observables are expected at detector level, but still semi-exclusive topologies has higher mean values. Jets are reconstructed using an anti-$k_T$ algorithm with a cone size of $\Delta R < 0.4$ where $R$ is the radius of the jet cone, ensuring well-isolated jet objects and reducing contamination from nearby hadronic activity. Forward hadronic activity was assessed using calorimeter towers in the hadronic forward HF region. The HF calorimeter provides coverage in the pseudorapidity range $3 < |\eta| < 5$, allowing for efficient tagging of forward energy deposits.\\ 

\begin{table}[h!]
\centering
\resizebox{1.0\columnwidth}{!}{%

\begin{tabular}{@{}l@{\hspace{2cm}}l@{}}
\toprule
\textbf{Object} & \textbf{Event selection criteria} \\
\midrule
Jets &
\begin{tabular}[t]{@{}l@{}}
\quad$\bullet$ $p_T > 25$ GeV \\
\quad$\bullet$ $|\eta| < 2.4$ \\
\quad$\bullet$ $\Delta R < 0.4$
\end{tabular} \\
\midrule
Jet multiplicity &
\begin{tabular}[t]{@{}l@{}}
\quad$\bullet$ $\geq 2$ jets \\
\quad$\bullet$ $\geq 1$ $b$-tagged jets \\
\quad$\bullet$ $\geq 1$ light-flavor jets
\end{tabular} \\
\midrule
Lepton multiplicity &
\begin{tabular}[t]{@{}l@{}}
\quad$\bullet$ Exactly one lepton 
\end{tabular} \\
\midrule
Electron &
\begin{tabular}[t]{@{}l@{}}
\quad$\bullet$ $p_T > 15$ GeV \\
\quad$\bullet$ $|\eta| < 2.4$
\end{tabular} \\
\midrule
Muon &
\begin{tabular}[t]{@{}l@{}}
\quad$\bullet$ $p_T > 10$ GeV \\
\quad$\bullet$ $|\eta| < 2.4$
\end{tabular} \\
\midrule
Charged tracks &
\begin{tabular}[t]{@{}l@{}}
\quad$\bullet$ $p_T > 0.4$ GeV \\
\quad$\bullet$ $|\eta| < 2.4$
\end{tabular} \\
\midrule
MET &
\begin{tabular}[t]{@{}l@{}}
\quad$\bullet$ $E^{\text{miss}}_{T} > 20$ GeV
\end{tabular} \\
\bottomrule
\end{tabular}
}
\caption{ \justifying Summary of detector-level selection criteria.}
\label{tab:detector_selection}
\end{table}

\noindent After the full reconstruction and selection procedure had been applied, event samples of comparable size were obtained for the three $t\bar{t}$ production mechanisms. Notably, the photon- and pomeron-induced processes were retained with higher efficiencies than the non-peripheral $t\bar{t}X$ contribution. In addition to this non-peripheral $t\bar{t}X$ background, further processes such as $W$+jets and Drell–Yan were also considered, confirming that the selection criteria had effectively suppressed backgrounds while preserving a substantial signal yield. The resulting efficiencies are summarized in Table~\ref{tab:efficiency_summary}. \\

\begin{table}[h!]
\centering
\resizebox{1.0\columnwidth}{!}{%
\begin{tabular}{l@{\hspace{3cm}}c}
\toprule
\textbf{Sample} & $\varepsilon_m \pm \Delta \varepsilon_m\ [\%]$ \\
\midrule
$p'\gamma p \rightarrow p't\bar{t}X$   & $16.95 \pm 0.04$ \\
$p'\mathbb{P}p \rightarrow p't\bar{t}X$ & $16.74 \pm 0.04$ \\
Non-peripheral $pp \rightarrow t\bar{t}X$        & $16.73 \pm 0.04$ \\
$p'\gamma p \rightarrow p'tWX$             & $16.95 \pm 0.03$ \\
$W + 1j$                                   & $0.18 \pm 0.01$ \\
$W + 2j$                                   & $0.55 \pm 0.02$ \\
$W + 3j$                                   & $1.00 \pm 0.04$ \\
Drell–Yan                                  & $0.08 \pm 0.01$ \\
\bottomrule
\end{tabular}
}
\caption{\justifying Selection efficiency $\varepsilon_m$ with statistical uncertainties $\Delta \varepsilon_m$ after full detector‐level $t\bar{t}$ reconstruction for each production mechanism.}
\label{tab:efficiency_summary}
\end{table}

\noindent To improve the ability to discriminate between semi-exclusive topologies and non-peripheral $t\bar{t}X$ production mechanism, a set of observables was selected based on their KS differences in variable distributions. These variables are designed to capture variations in event multiplicities and energy depositions, particularly in the forward region of the detector. The variables analyzed include:

\begin{itemize}
  \item \textbf{$N_\mathrm{jets}$}: Total number of jets with $p_T > 25~\mathrm{GeV}$ and $|\eta| < 2.4$.

  \item \textbf{$N_\mathrm{ch}$}: Number of stable charged particles (tracks) with $p_T > 0.4~\mathrm{GeV}$ and $|\eta| < 2.4$.

  \item \textbf{$N^\mathrm{fwd}_\mathrm{jets}$}: Number of jets in the forward region, defined by $|\eta| > 3.0$.

  \item \textbf{FRG}: Forward rapidity gap, defined as the minimum pseudorapidity distance to $\eta = +2.4$ from any track.

  \item \textbf{HF (-)}: Sum of energy in the backward hadronic forward (HF) calorimeter region, $-5.0 \leq \eta \leq -3.0$.

  \item \textbf{Min HF tower $E$}: Minimum energy among all calorimeter towers within $3.0 \leq \eta \leq 5.0$ and $-5.0 \leq \eta \leq -3.0$.

  \item \textbf{Max HF tower $E$}: Maximum energy among all calorimeter towers within $3.0 \leq \eta \leq 5.0$ and $-5.0 \leq \eta \leq -3.0$.

  \item \textbf{HF$_\mathrm{calo}^{\max}$ min.}: Minimum of the two maximum tower energies in the HF$+$ and HF$-$ hemispheres:
  \[
  \text{HF}_\mathrm{calo}^{\max}\ \text{min.} = \min(E^{\max}_\mathrm{HF(+)} , E^{\max}_\mathrm{HF(–)}).
  \]

  \item \textbf{HF min}: Minimum of the total HF$+$ and HF$-$ energy sums:
  \[
  \text{HF min} = \min\left(\sum E^\text{HF(+)} , \sum E^\text{HF(–)}\right).
  \]
\end{itemize}

\begin{figure*}[!ht]
    \centering
    \begin{subfigure}[t]{0.32\textwidth}
        \includegraphics[width=\linewidth]{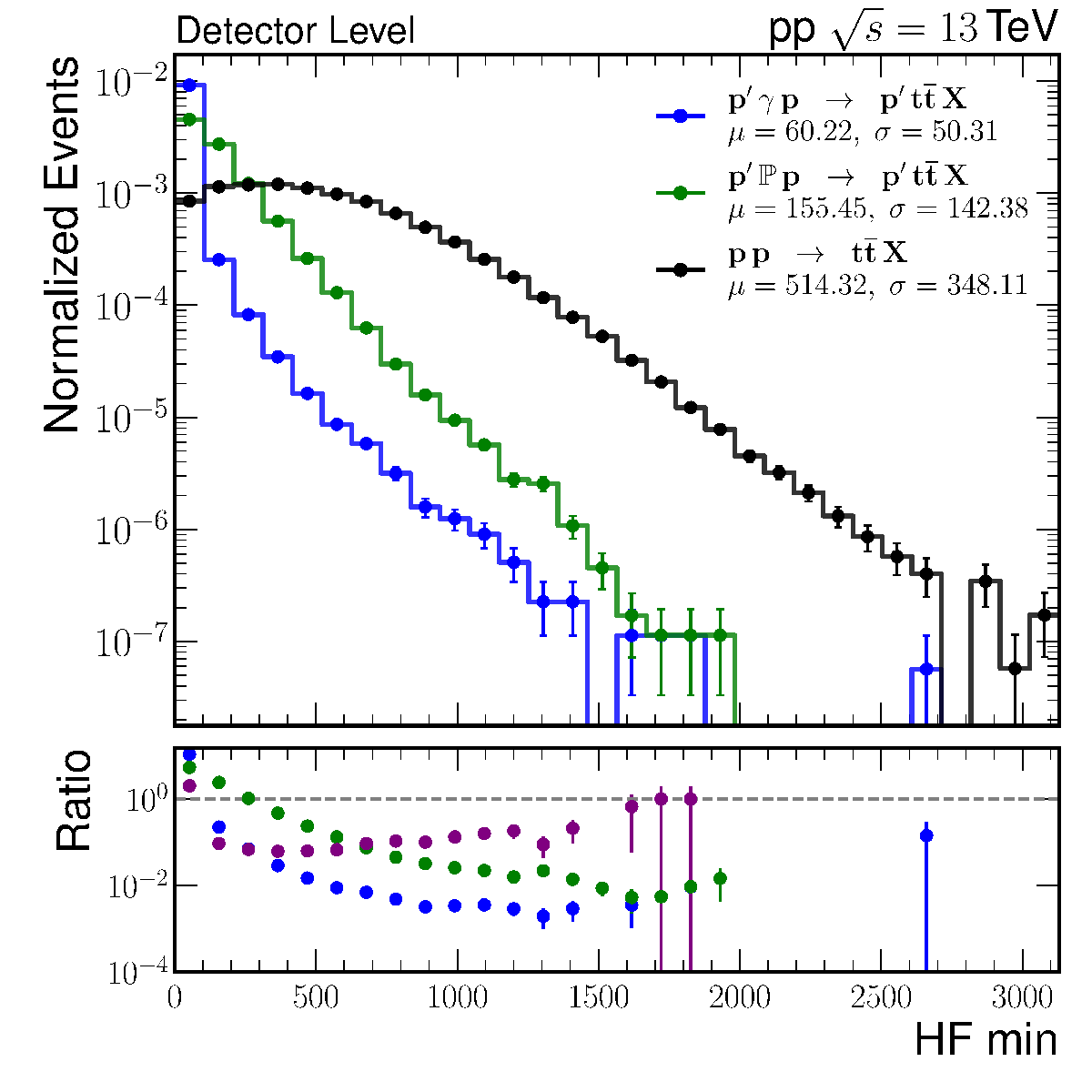}
    \end{subfigure}\hfill
    \begin{subfigure}[t]{0.32\textwidth}
        \includegraphics[width=\linewidth]{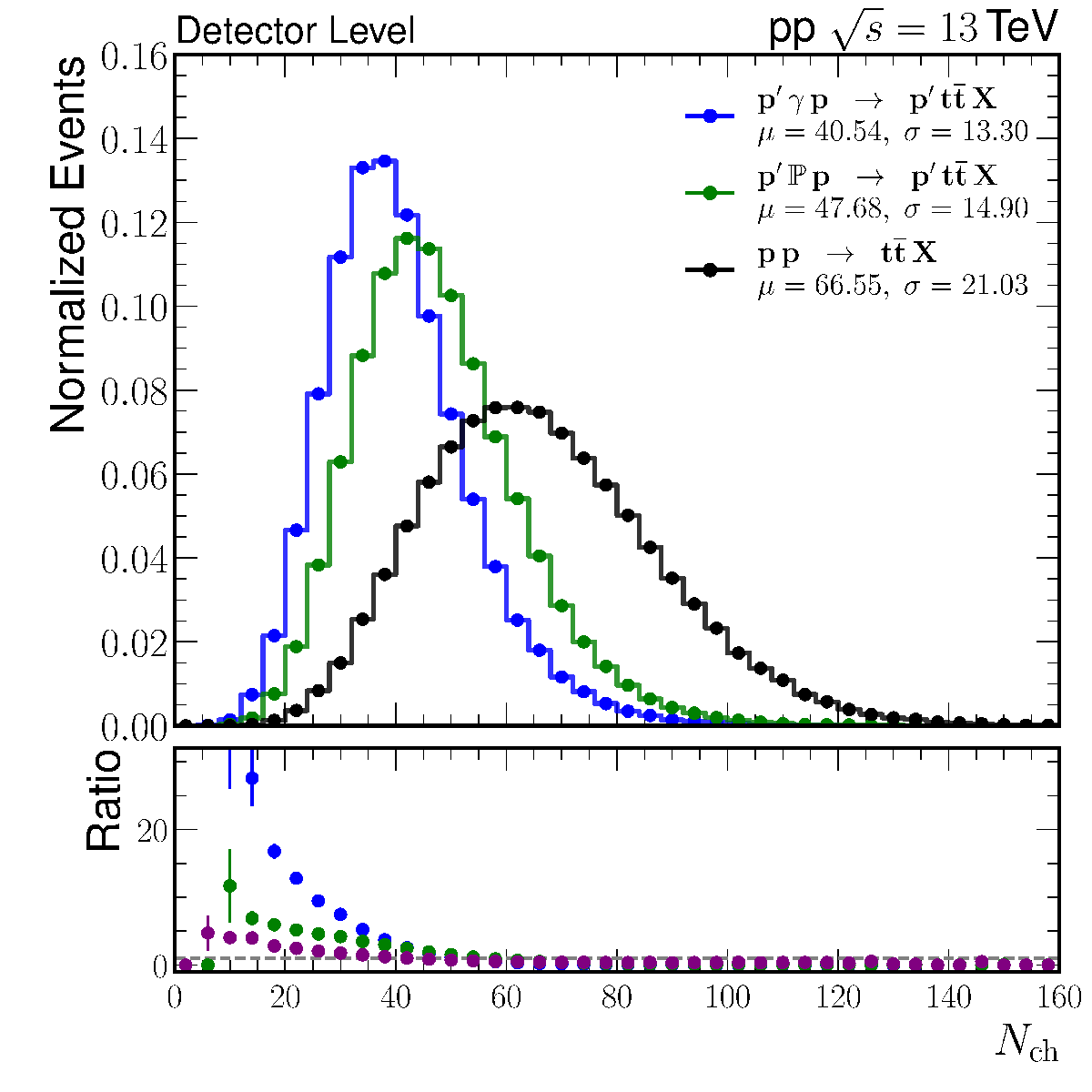}
    \end{subfigure}\hfill
    \begin{subfigure}[t]{0.32\textwidth}
        \includegraphics[width=\linewidth]{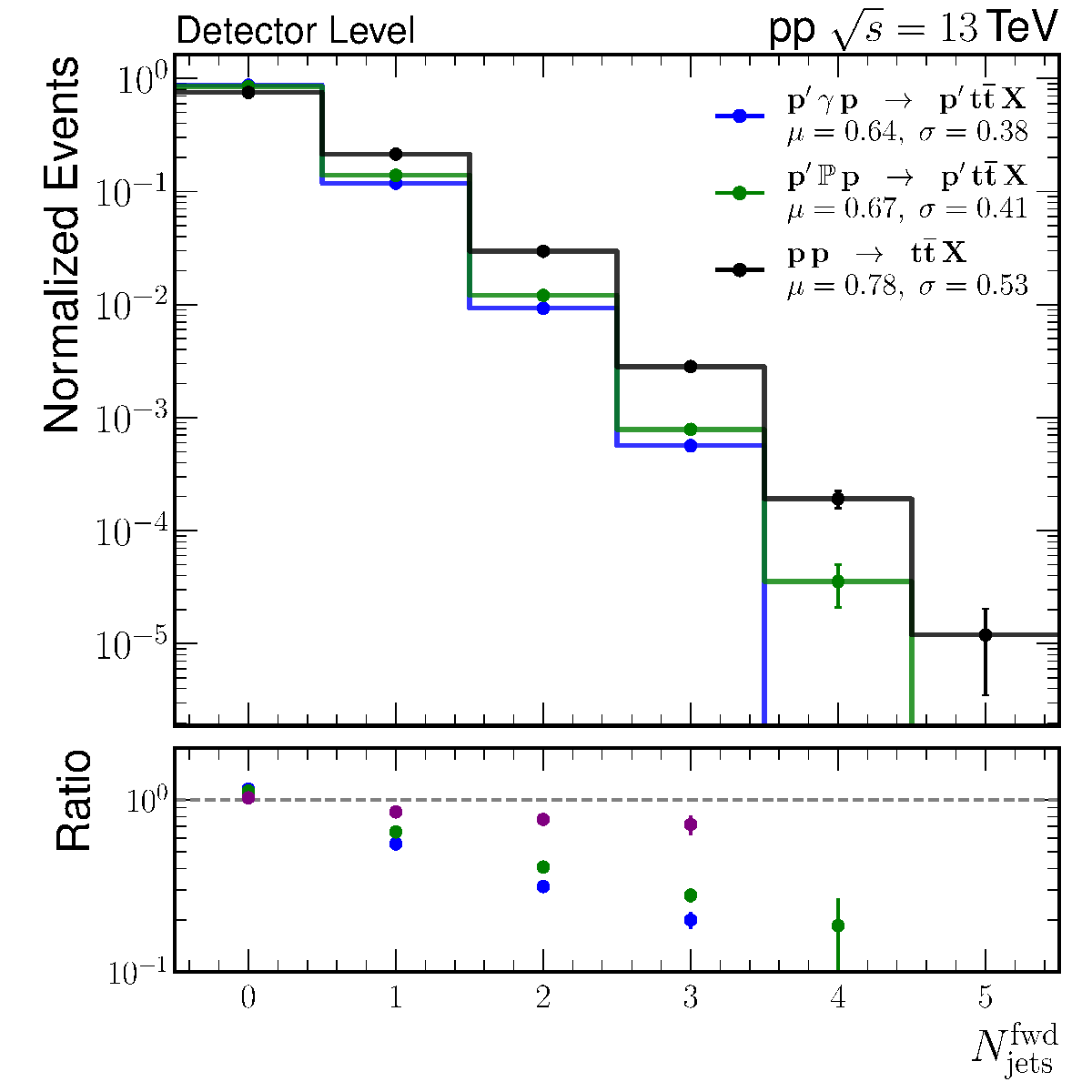}
    \end{subfigure}

    \vspace{1em}

    \begin{subfigure}[t]{0.32\textwidth}
        \includegraphics[width=\linewidth]{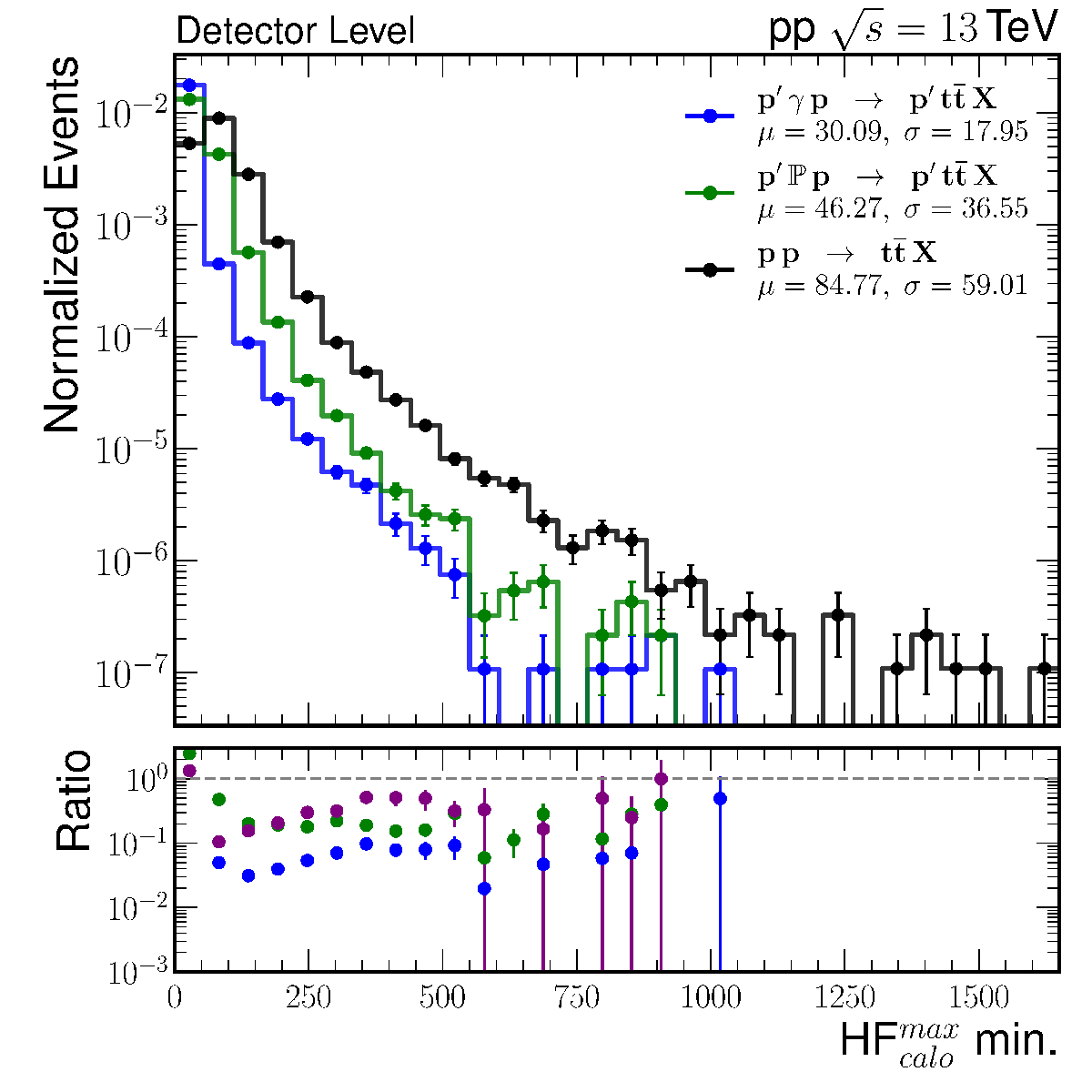}
    \end{subfigure}\hfill
    \begin{subfigure}[t]{0.32\textwidth}
        \includegraphics[width=\linewidth]{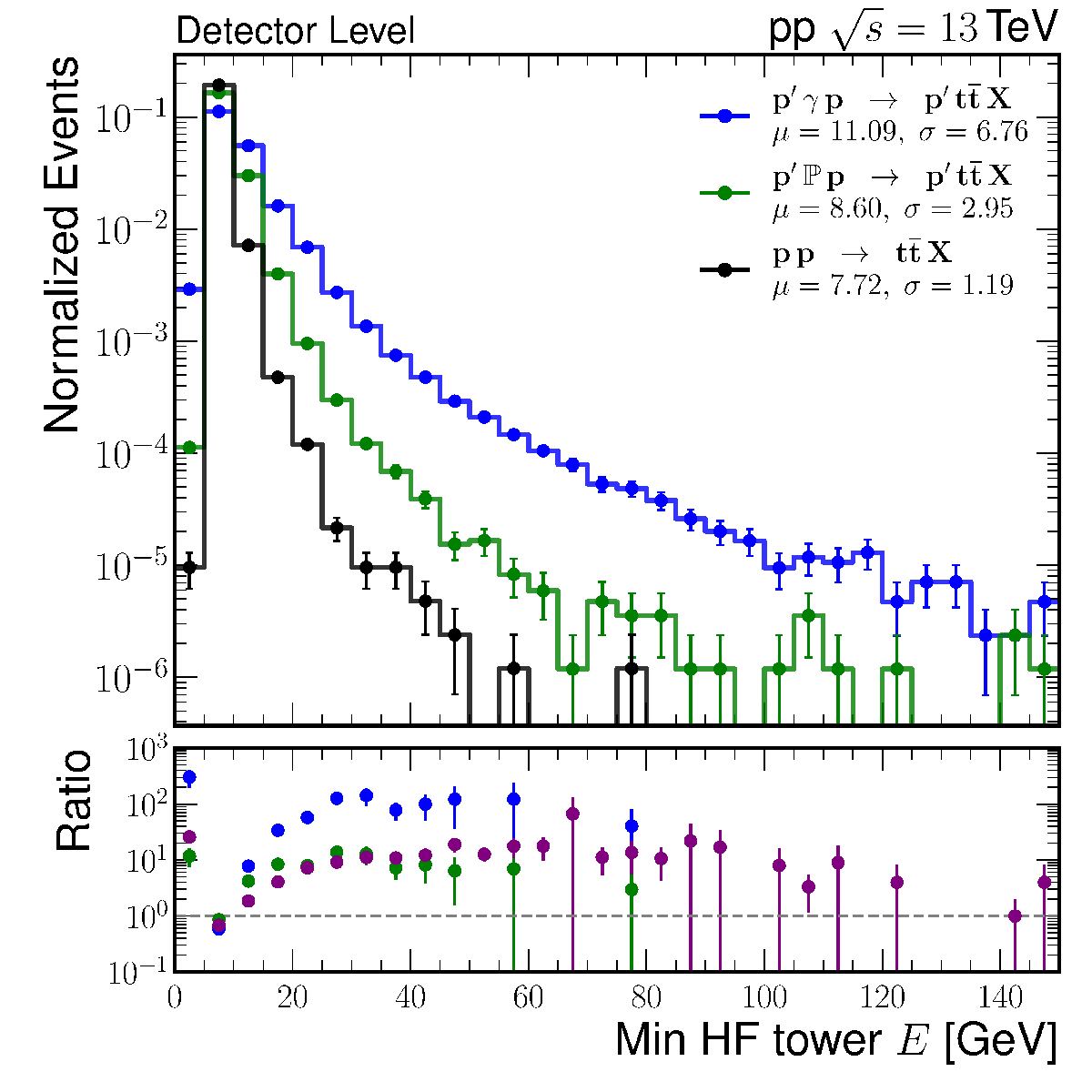}
    \end{subfigure}\hfill
    \begin{subfigure}[t]{0.32\textwidth}
        \includegraphics[width=\linewidth]{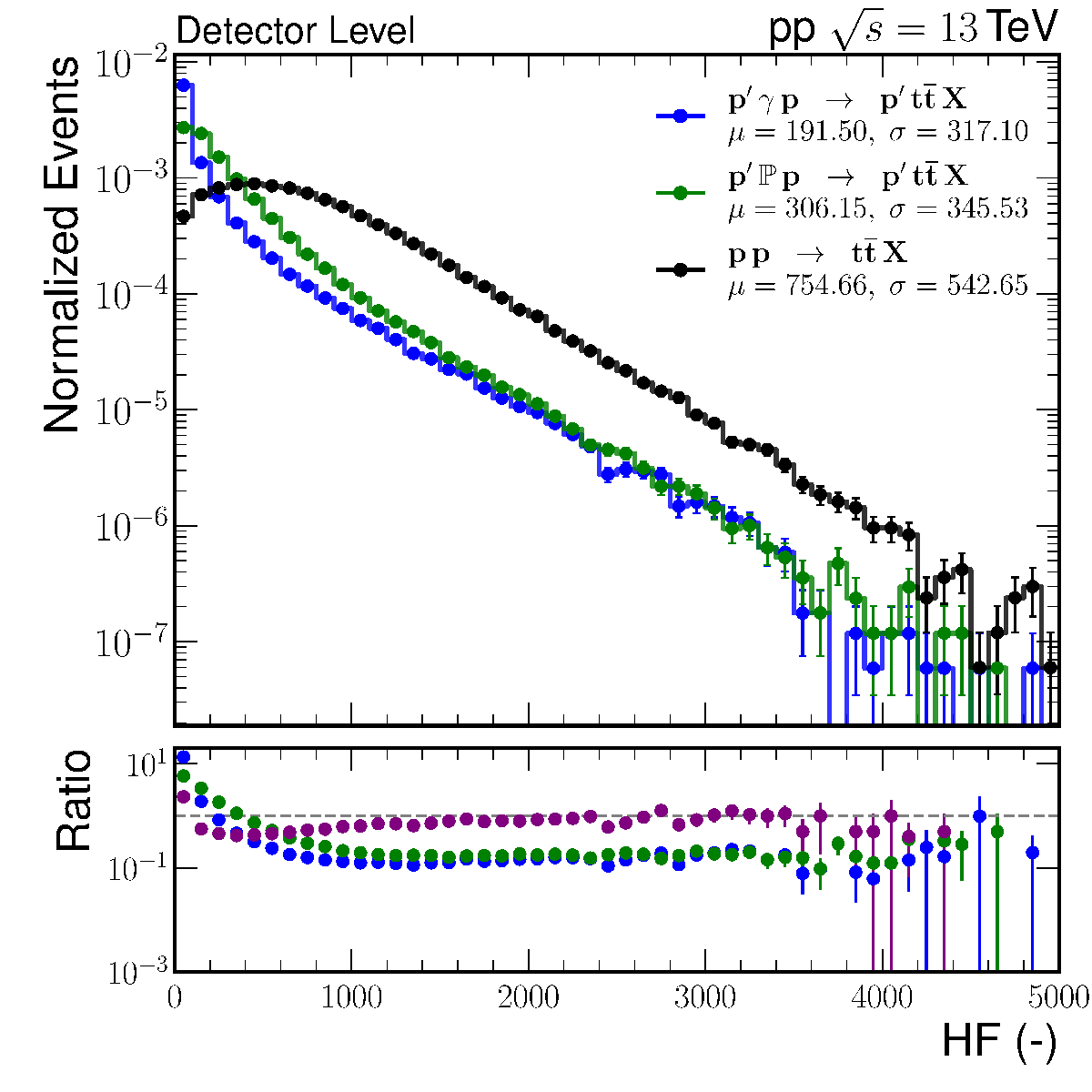}
    \end{subfigure}

    \vspace{1em}

    \begin{subfigure}[t]{0.32\textwidth}
        \includegraphics[width=\linewidth]{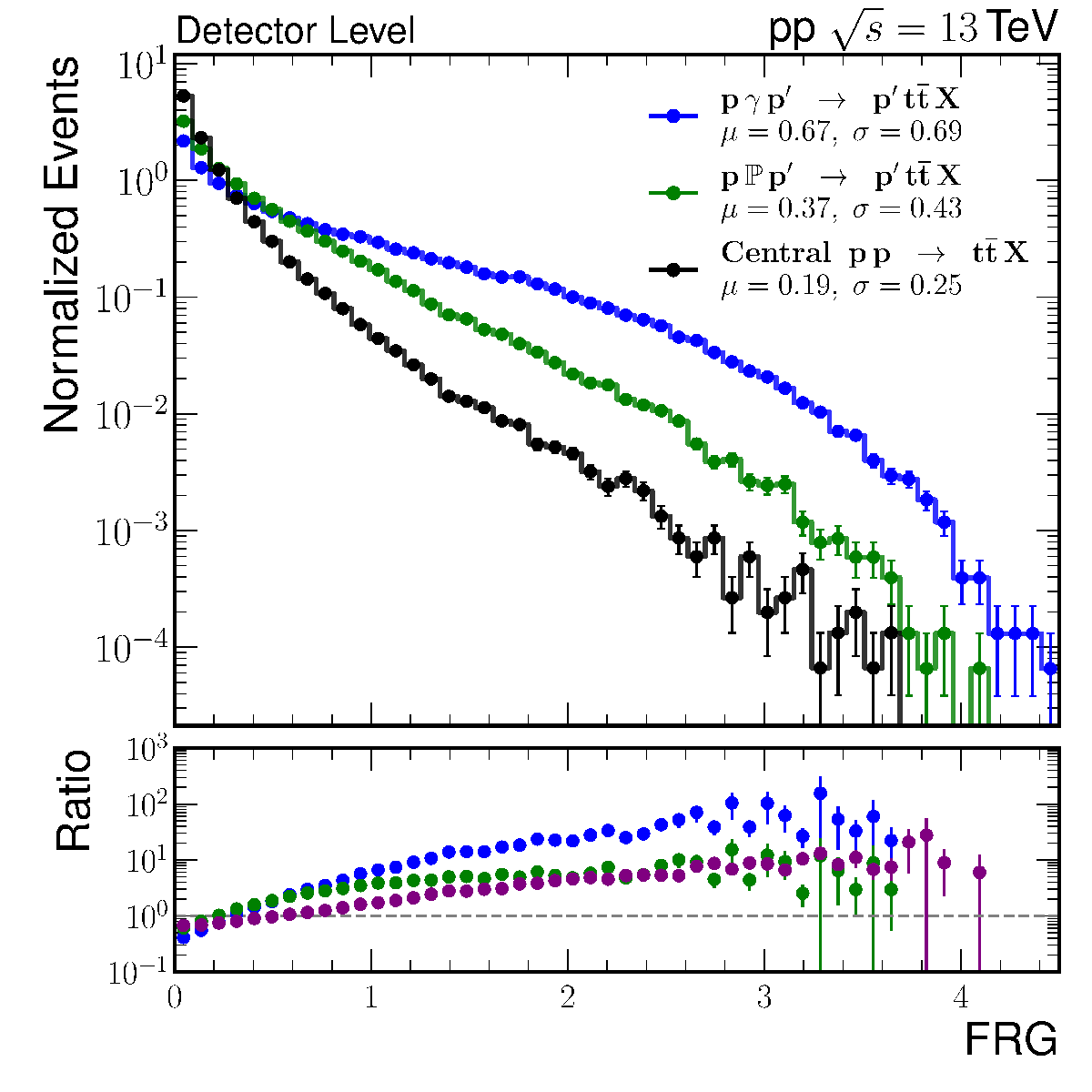}
    \end{subfigure}\hfill
    \begin{subfigure}[t]{0.32\textwidth}
        \includegraphics[width=\linewidth]{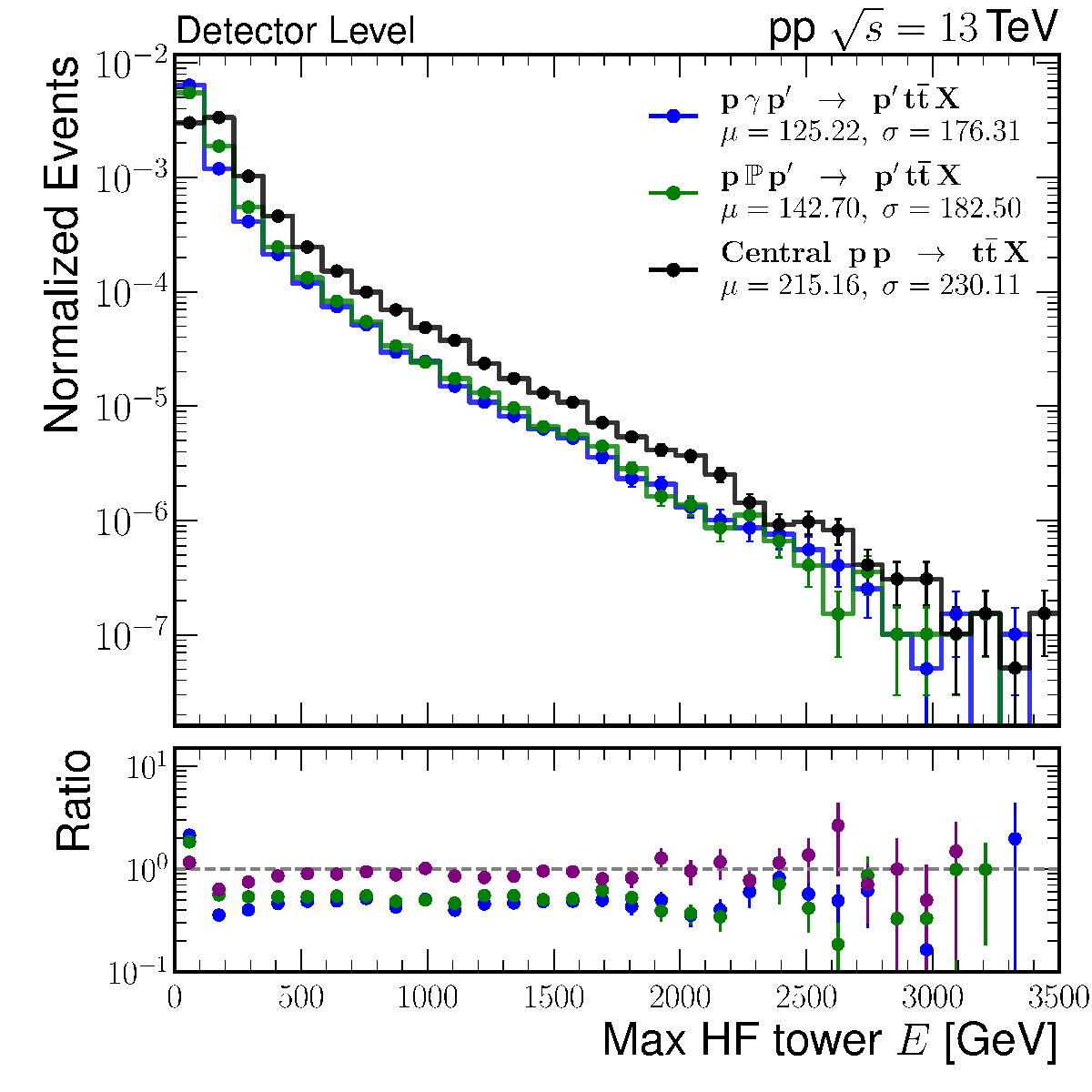}
    \end{subfigure}\hfill
    \begin{subfigure}[t]{0.32\textwidth}
        \includegraphics[width=\linewidth]{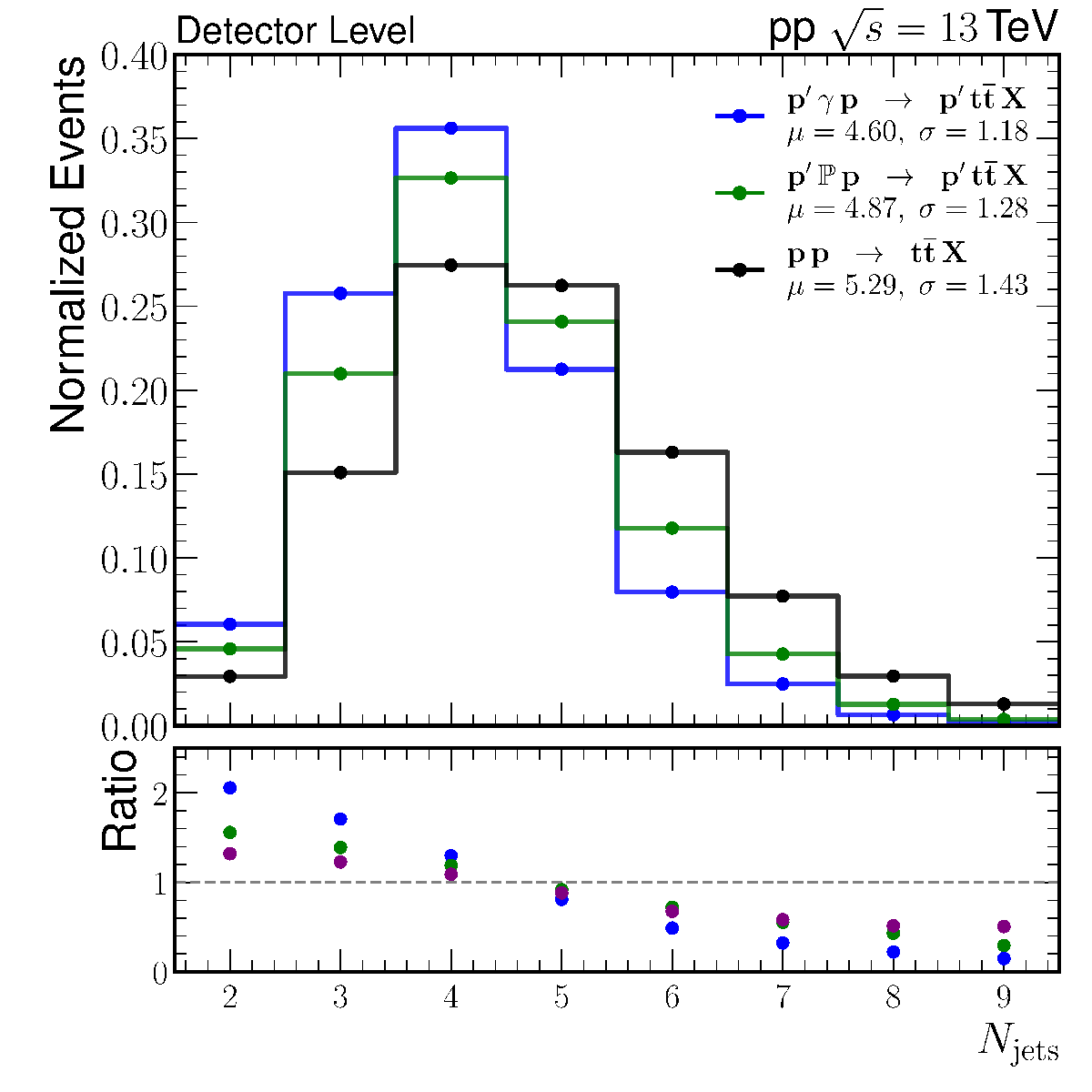}
    \end{subfigure}

    \caption{\justifying Nine detector-level distributions with the highest KS separation power between pomeron or photon induced and non-peripheral $t\bar{t}X$ events.}

    \label{fig:ks_top9}
\end{figure*}

\noindent The distributions of the nine observables for the three simulated samples ($p'\gamma p \rightarrow p' t\bar{t}X$, $p'\mathbb{P}p \rightarrow p' t\bar{t}X$, and $pp \rightarrow t\bar{t}X$) are shown in Figure~\ref{fig:ks_top9}. Visible and quantified differences are spotted in several observables, particularly those sensitive to forward activity and charged particle multiplicity, motivating their use as input features for production-mode classification. To quantitatively assess their discriminating power, KS test statistics were computed for each observable, comparing all three production mode pairs with each other. The best variables are summarized in Table~\ref{tab:ks_summary}, with uncertainties obtained from the standard deviation of $10^3$ bootstrap resamples of each distribution pair. Among all observables, \textbf{HF min} achieved the highest KS score in all three comparisons: $p'\gamma p\rightarrow p' t\bar{t}X$ vs $p'\mathbb{P}p \rightarrow p' t\bar{t}X$, $p'\gamma p$ vs $pp \rightarrow t\bar{t}X$, and $p'\mathbb{P}p \rightarrow p' t\bar{t}X$ vs $pp \rightarrow t\bar{t}$. This indicates that forward calorimetric energy imbalance is a strong discriminator and the most sensitive observable between semi-exclusive and non-peripheral $t\bar{t}X$ production mechanisms. All selected observables collectively serve as a solid set of variables for a multivariate analysis aimed at improving classification performance across the three production mechanisms.\\

\begin{table}[h!]
\centering
\resizebox{1.0\columnwidth}{!}{%
\begin{tabular}{lccc}
\toprule
\textbf{Observable} 
& \multicolumn{3}{c}{\textbf{KS values}} \\
\cmidrule(lr){2-4}
& $(p'\gamma p ,\, p'\mathbb{P}p)$ 
& $(p'\gamma p ,\, pp )$ 
& $(p'\mathbb{P}p ,\, pp)$ \\
\midrule
\textbf{$N_\mathrm{jets}$}                         & 0.092 ± 0.002 & 0.220 ± 0.002 & 0.128 ± 0.002 \\
\midrule
\textbf{$N_\mathrm{ch}$}                           & 0.216 ± 0.002 & 0.568 ± 0.001 & 0.410 ± 0.002 \\
\midrule
\textbf{$N^\mathrm{fwd}_\mathrm{jets}$}            & 0.024 ± 0.001 & 0.118 ± 0.001 & 0.094 ± 0.001 \\
\midrule
\textbf{FRG}                                       & 0.201 ± 0.002 & 0.401 ± 0.001 & 0.232 ± 0.002 \\
\midrule
\textbf{HF (-)}                                    & 0.400 ± 0.001 & 0.649 ± 0.001 & 0.476 ± 0.002 \\
\midrule
\makecell[l]{\textbf{Min HF} \\ \textbf{tower $E$}} & 0.257 ± 0.002 & 0.497 ± 0.001 & 0.278 ± 0.002 \\
\midrule
\makecell[l]{\textbf{Max HF} \\ \textbf{tower $E$}} & 0.234 ± 0.001 & 0.478 ± 0.001 & 0.310 ± 0.002 \\
\midrule
\makecell[l]{\textbf{HF$_\mathrm{calo}^{\max}$} \\ \textbf{min.}} & 0.746 ± 0.001 & 0.843 ± 0.001 & 0.431 ± 0.002 \\
\midrule
\makecell[l]{\textbf{HF} \\ \textbf{min}}          & {\textbf{0.752 ± 0.001}} & \textbf{0.880 ± 0.001} & \textbf{0.568 ± 0.001} \\
\bottomrule
\end{tabular}%
}
\caption{\justifying 
Kolmogorov–Smirnov test statistics with bootstrap uncertainties for each observable, comparing $p'\gamma p \rightarrow p' t\bar{t}X$, $p'\mathbb{P}p \rightarrow p' t\bar{t}X$, and $pp \rightarrow t\bar{t}X$ production modes. The highest KS value in each column is shown in bold.
}
\label{tab:ks_summary}
\end{table}

\noindent Previous studies have established the connection between forward rapidity gaps and diffractive dynamics, as well as the suppressed particle multiplicities characteristic of color-singlet exchanges in both diffractive and photon-induced processes~\cite{SciPostPhysProc.8.081, CMS_Collaboration_2024, atlas_2012_rapidity_gap}. Moreover, measurements of forward energy flow in the HF region have proven instrumental for identifying diffractive topologies in hadronic collisions~\cite{vanhaevermaet2011}. In CMS the charged-particle multiplicity in $\gamma p$ interactions was found to be significantly lower than in hadronic $pPb$ collisions for comparable event classes, and the mean track multiplicity $\langle N_{\text{trk}}^{\text{offline}}\rangle$ in $\gamma p$ events was measured to remain below $\sim 3$ even in the highest multiplicity bin considered~\cite{Tumasyan_2023_}. In another study, forward–rapidity–gap selected $pPb$ events at $\sqrt{s_{NN}}=8.16~\text{TeV}$ have been shown to exhibit low charged–particle multiplicities, with the mean track count dropping by over 50\% as the gap widens and the distributions remaining below $N_{\mathrm{trk}}^{\mathrm{offline}}\!\sim\!50$, consistent with expectations for color–singlet exchange. Our results follow this established low-multiplicity behaviour for diffractive topologies and extend it to high-mass $t\bar t$ final states~\cite{CMS-PAS-HIN-22-004}. However, no studies have specifically analyzed top-quark pair production in a semi-exclusive configuration, offering a new perspective on the interplay between color-neutral exchange mechanisms in $t\bar{t}$ production and the corresponding separation power of KS values of detector-level observables with respect to the background. Furthermore, based on these detector level observables, neural networks and likelihood scans were introduced to quantify the expected signal strengths and to estimate the integrated luminosity required to achieve a $5\sigma$ observation, as detailed in Sections~6 and~7.


\section{\label{sec:level5} Neural network classification} %

\begin{figure*}[t]
    \centering
    \begin{subfigure}{0.32\textwidth}
        \includegraphics[width=\linewidth]{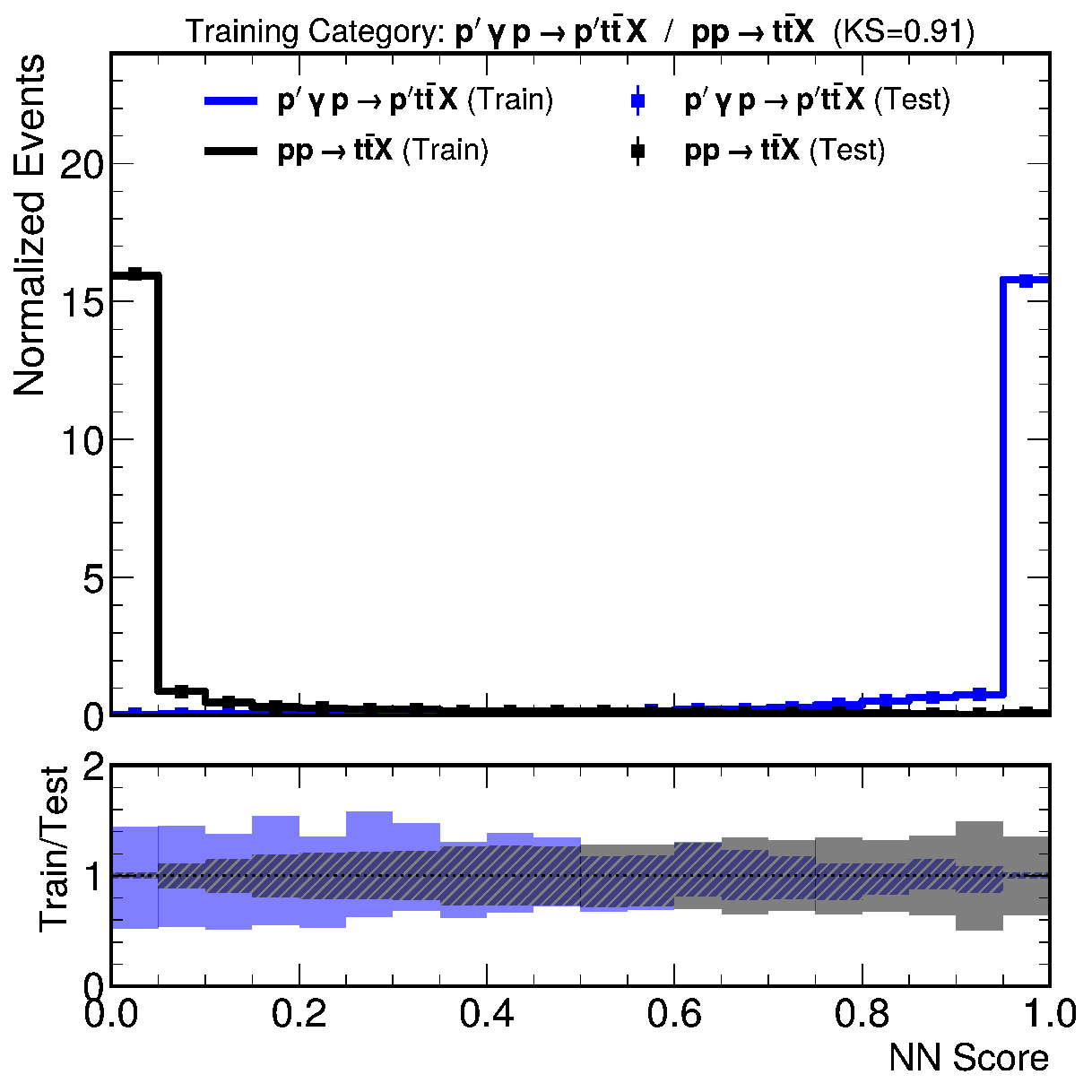}
        \caption{}
    \end{subfigure}
    \begin{subfigure}{0.32\textwidth}
        \includegraphics[width=\linewidth]{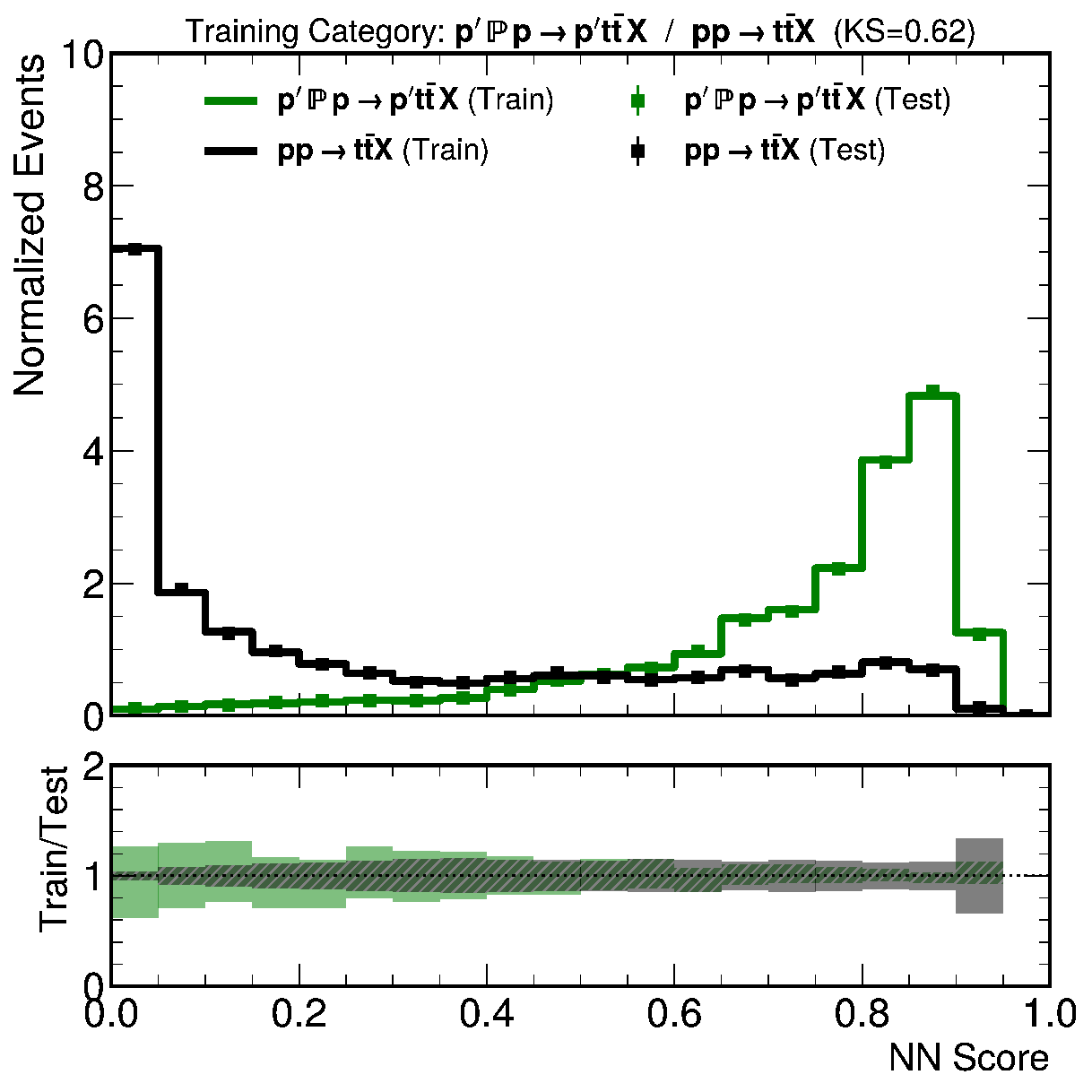}
        \caption{}
    \end{subfigure}
    \begin{subfigure}{0.32\textwidth}
        \includegraphics[width=\linewidth]{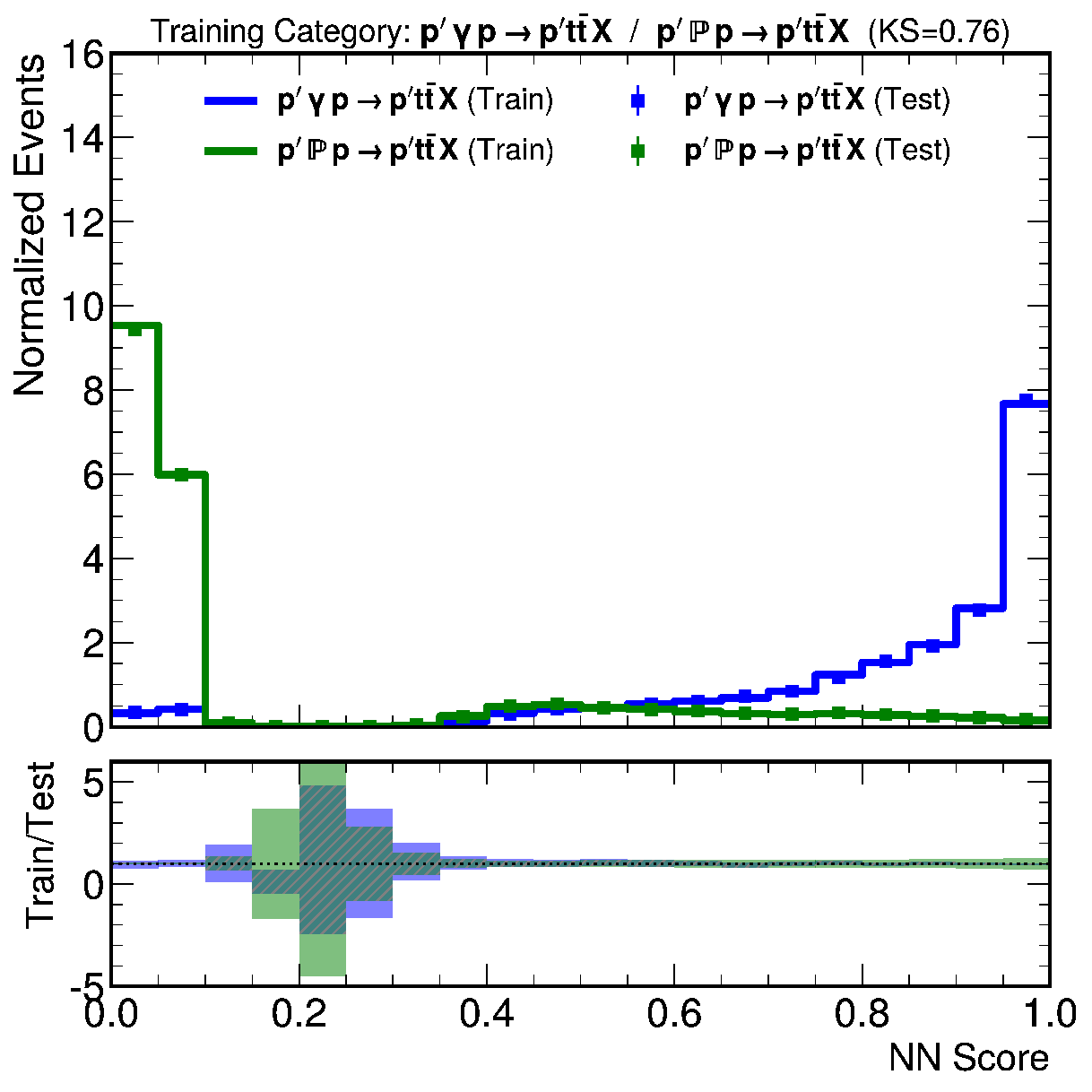}
        \caption{}
    \end{subfigure}
    \caption{\justifying Neural network score distributions for the classification of 
    (a) $p^\prime\,\gamma\,p \rightarrow p^\prime\,t\bar{t}\,X$ vs.\ $pp \rightarrow t\bar{t}\,X$, 
    (b) $p^\prime\,\mathbb{P}\,p \rightarrow p^\prime\,t\bar{t}\,X$ vs.\ $pp \rightarrow t\bar{t}\,X$, 
    and (c) $p^\prime\,\gamma\,p \rightarrow p^\prime\,t\bar{t}\,X$ vs.\ $p^\prime\,\mathbb{P}\,p \rightarrow p^\prime\,t\bar{t}\,X$. The largest KS separation is observed for the photon-induced process compared to the non-peripheral $t\bar{t}X$ production (a).}
    \label{fig:score_distributions_tensorflow}
\end{figure*}

\begin{figure}[!ht]
    \centering
    \includegraphics[width=\columnwidth]{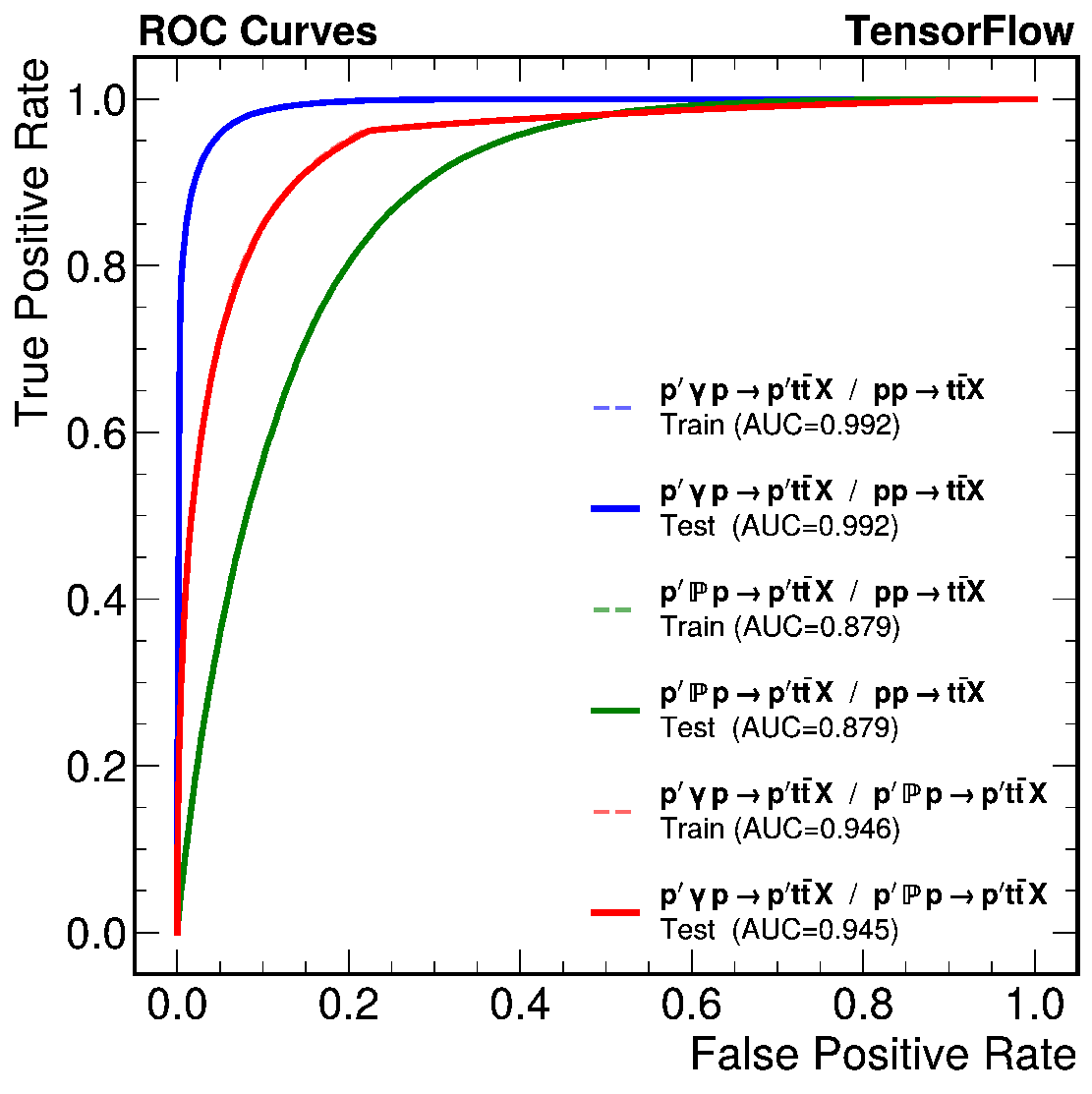}
    \caption{ROC curves for the three binary classifiers categories.}
    \label{fig:roc_all_models}
\end{figure}

\noindent
The selected observables serve as input to a fully multivariate analysis. A feed-forward neural network was implemented using \texttt{TensorFlow} and \texttt{Keras}~\cite{tensorflow2015-whitepaper}. Unlike low-dimensional cut-based methods, the neural network models nonlinear correlations and complex decision boundaries based on the nine input variables opposed to applying further cuts over them and then allowing for a stronger separation power. The network was trained on labeled photon-, pomeron-, and non-peripheral $t\bar{t}X$ samples. It learned optimal decision boundaries in the high-dimensional feature space by maximizing the area under the Receiving Operational Characteristic curve (ROC curve) and reducing misclassifications rates across all event classes. The classifier outputs distributions, expressed as normalized neural network scores (NN scores), are shown in Figure~\ref{fig:score_distributions_tensorflow}. Clear separation is observed in the output distributions across all three binary training categories. The lower panels show excellent agreement between training and test sets, indicating no signs of overtraining occurred. The best separation assessed via KS test whose larger output (closer to 1.0) indicates great discrimination between two distributions is observed in the classification between $p'\gamma p$ and non-peripheral $t\bar{t}X$ events, with a KS statistic of 0.91, followed by the $p'\gamma p$ vs.\ $p'\mathbb{P}p$ comparison with a KS value of 0.76. $p'\mathbb{P}p$ vs.\ non-peripheral $t\bar{t}X$ classification resulted in a KS score of 0.62. This hierarchy reflects the physical similarity between photon- and pomeron-induced processes, both of which exhibit semi-exclusive topologies characterized by reduced non-peripheral $t\bar{t}X$ activity and forward signatures. As both are dominated by color-singlet exchange, their shared kinematic structure limits the network’s ability to fully disentangle them. In contrast, non-peripheral $t\bar{t}X$ production differs more substantially in event topology, especially with respect photon-induced case, yielding higher separation power. Table~\ref{tab:ai_performance} presents the full set of performance metrics including AUC, accuracy, precision, recall, F1 score, specificity, and KS for each training category. The AUC values, in particular, mirror the KS hierarchy observed in the score distributions: the $p'\gamma p$ vs.\ non-peripheral $t\bar{t}X$ production achieves an AUC of 0.992, the $p'\gamma p$ vs.\ $p'\mathbb{P}p$ production reaches 0.945, and the $p'\mathbb{P}p$ vs.\ non-peripheral $t\bar{t}X$ production trails with 0.879. These results are further supported by the ROC curves shown in Figure~\ref{fig:roc_all_models}, which illustrate the trade-off between true and false positive rates for each classifier. In all cases, the close agreement between training and test AUCs confirmed the network’s robustness and generalization performance, with no indication of overtraining.

\begin{table}[!htbp]
\centering
\resizebox{1.0\columnwidth}{!}{%
\begin{tabular}{@{}llcc@{}}
\toprule
\textbf{Training category} & \textbf{Metric}        & \textbf{Train} & \textbf{Test} \\
\midrule
($p'\gamma p\rightarrow p' t\bar{t}X$, \,$pp\rightarrow t\bar{t}X$)
    & AUC             & 0.9917 & 0.9917 \\
    & Accuracy        & 0.9536 & 0.9542 \\
    & Precision       & 0.9468 & 0.9475 \\
    & Recall          & 0.9612 & 0.9617 \\
    & F\(\!1\) Score  & 0.9539 & 0.9545 \\
    & KS Statistic    & 0.9089 & 0.9089 \\
    & Specificity     & 0.9467 & 0.9467 \\
\midrule
($p'\mathbb{P}p\rightarrow p' t\bar{t}X$, \,$pp\rightarrow t\bar{t}X$)
    & AUC             & 0.8783 & 0.8785 \\
    & Accuracy        & 0.8069 & 0.8078 \\
    & Precision       & 0.7695 & 0.7713 \\
    & Recall          & 0.8762 & 0.8752 \\
    & F\(\!1\) Score  & 0.8194 & 0.8200 \\
    & KS Statistic    & 0.6162 & 0.6162 \\
    & Specificity     & 0.7404 & 0.7404 \\
\midrule
($p'\gamma p\rightarrow p' t\bar{t}X$, \,$p'\mathbb{P}p\rightarrow p' t\bar{t}X$)
    & AUC             & 0.9466 & 0.9455 \\
    & Accuracy        & 0.8830 & 0.8816 \\
    & Precision       & 0.8962 & 0.8963 \\
    & Recall          & 0.8664 & 0.8630 \\
    & F\(\!1\) Score  & 0.8810 & 0.8793 \\
    & KS Statistic    & 0.7641 & 0.7641 \\
    & Specificity     & 0.9002 & 0.9002 \\
\bottomrule
\end{tabular}
}
\caption{\justifying Performance metrics of neural network classifiers for each pairwise training category, evaluated on both training and test datasets.  The event categories correspond to: $p'\gamma p\rightarrow p' t\bar{t}X$, $p'\mathbb{P}p\rightarrow p' t\bar{t}X$, and non-peripheral $pp \rightarrow t\bar{t}X$.}
\label{tab:ai_performance}
\end{table}

\noindent Altogether, these results shows the discriminating power of the selected input observables and the effectiveness of neural networks for modeling complex correlations between event classes. The strong classification achieved for $p'\gamma p$ vs.\ non-peripheral $t\bar{t}X$ and $p'\gamma p$ vs.\ $p'\mathbb{P}p$ suggests that topologies with prominent forward gaps or isolated forward energy are well-separated in the learned representation, and set a particular positive scenario towards enhancing particularly the photon-induced observation. However, $p'\mathbb{P}p$ and non-peripheral $t\bar{t}X$ events points towards a potential improvement. \\

\section{\label{sec:level5} Likelihood scan} %

\begin{figure*}[!ht]
  \centering
  \begin{subfigure}{0.32\textwidth}
    \includegraphics[width=\linewidth]{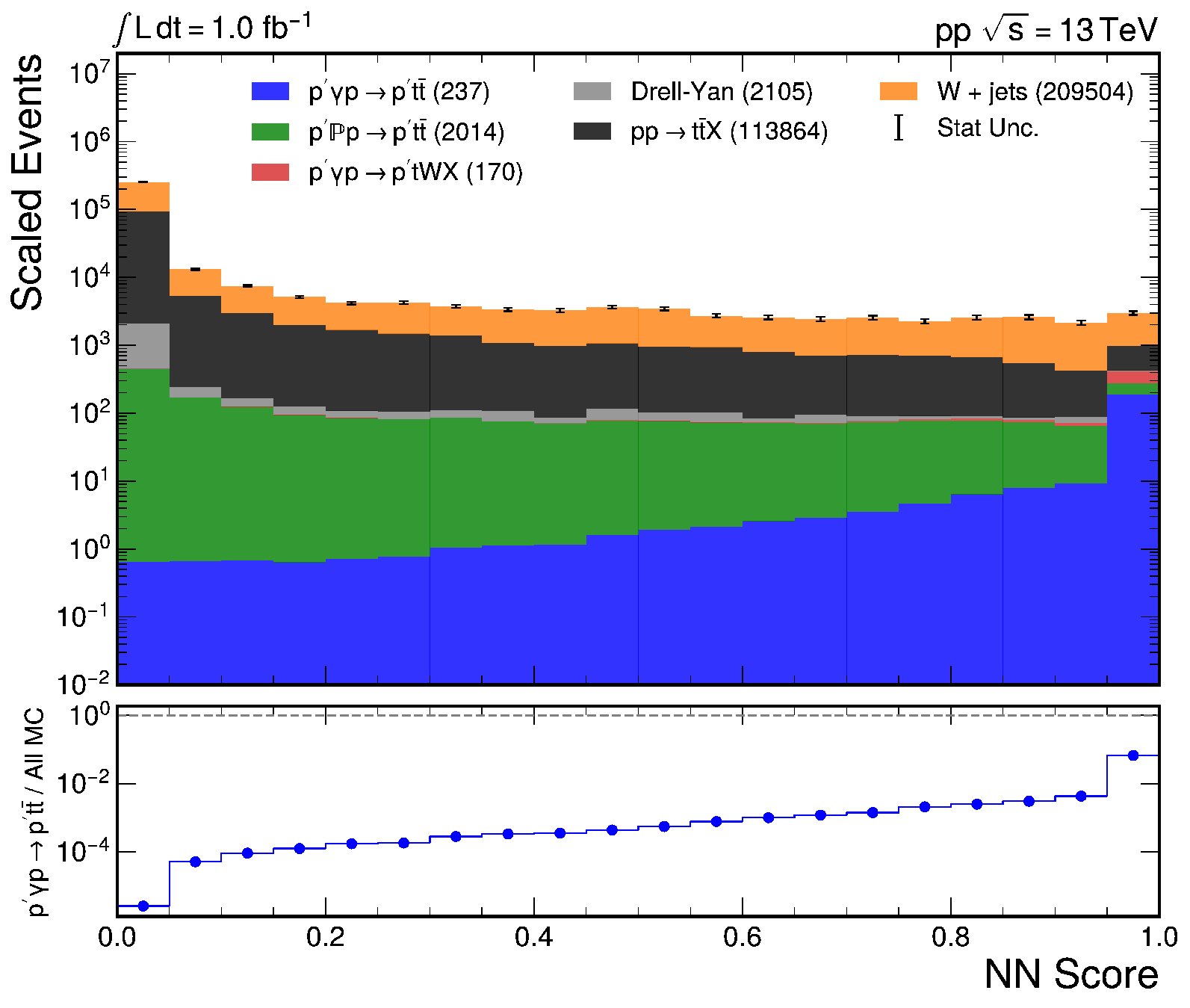}
    \caption{}
  \end{subfigure}
  \begin{subfigure}{0.32\textwidth}
    \includegraphics[width=\linewidth]{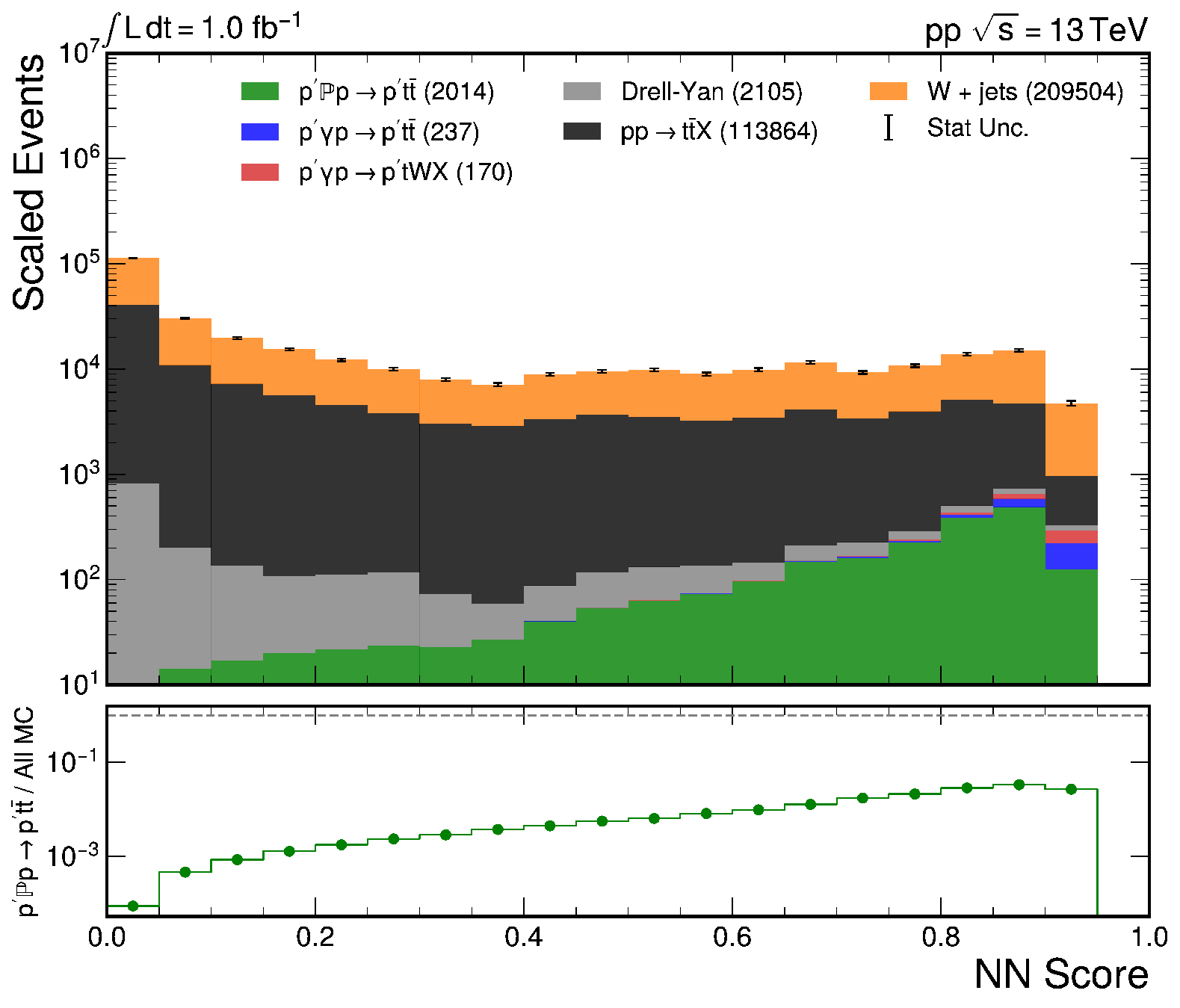}
    \caption{}
  \end{subfigure}
  \begin{subfigure}{0.32\textwidth}
    \includegraphics[width=\linewidth]{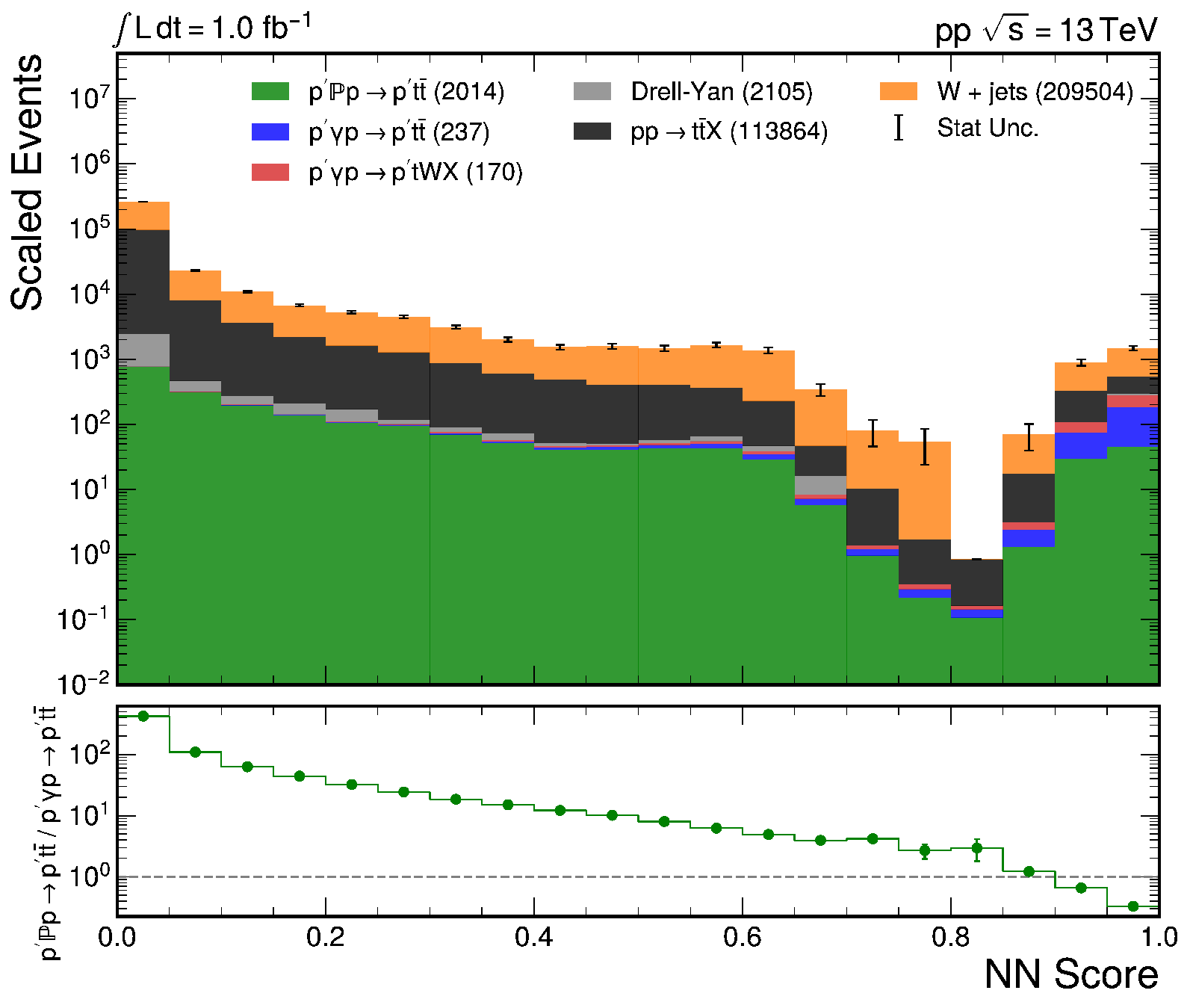}
    \caption{}
  \end{subfigure}
    \caption{Neural network score distributions used in the profile-likelihood fit:
    (a) $p'\gamma p\to \,p't\bar{t}X$ vs.\ $pp \to t\bar{t}X$,
    (b) $p'\mathbb{P}p\to t\bar{t}\,p'$ vs.\ $pp \to t\bar{t}X$,
    (c) $p'\gamma p\to t\bar{t}\,p'$ vs.\ $p'\mathbb{P}p\to \,p't\bar{t}X$.}
    \label{fig:score_distributions}
\end{figure*}

\noindent The neural-network classification stage produced integrated scores capable of discriminating between photon-induced, pomeron-induced and non-peripheral $t\bar{t}X$ events. These discriminants are used in the statistical inference stage to extract the level of sensitivity at which semi-exclusive processes could be observed at current and future LHC luminosities. This is done via the extraction of the signal strengths $\mu_{\mathbb{P}}$ and $\mu_{\mathbb{\gamma}}$ defined as $\mu_{j} = \frac{\sigma_{\text{j}}}{\sigma_{\text{SM}}}$, where $\sigma_{i}$ represents the Asimov dataset and $\mu_{i}$ its rate comparison with respect to SM expectation, $\mu$ $=$ 1 would correspond to a template that represents the SM prediction for the two semi-exclusive production modes. The statistical model, implemented within the \texttt{Combine} framework~\cite{combine}, employs a shape–based, binned profile–likelihood method, with template normalizations controlled by the respective $\mu$ parameters and statistical uncertainties treated using the Barlow–Beeston method~\cite{Barlow}. Using the three score distributions described in section VI, each corresponding to one of the binary neural–network trainings: photon–induced vs. pomeron–induced, pomeron–induced vs. non-peripheral $t\bar{t}X$ background, and photon–induced vs. non-peripheral $t\bar{t}X$ background. The predicted event yields for each process are first scaled to their SM expectations and a luminosity of $1\,\text{fb}^{-1}$, according to

\begin{equation}
N^{\text{expected}}_m = \varepsilon_m \, \sigma_m \, \mathcal{L},
\end{equation}
with $\sigma_m$ the process cross section and $\mathcal{L}$ the integrated luminosity. 
The neural–network outputs are then converted into scaled to SM discriminant histograms, which are shown in Figure~\ref{fig:score_distributions}. 
Additional background contributions from W$+$jets, Drell–Yan and single photo–induced top are added to the stacked histograms. 
These three distributions form the fundamental shape templates of the profile–likelihood model, serving as the direct inputs to \textsc{Combine} after a final filter is applied, aiming to discriminate the semi–exclusive processes between them. 
The use of Asimov expectations for normalisation follows a similar procedure published in~\cite{Elwood:2018qsr}. Maximum level of significance between photon-induced and pomeron-induced events is achieved by locating the NN score in the photon--pomeron classifier at the bottom of Figure~\ref{fig:score_distributions} (c) that maximizes the photon or pomeron Asimov significance~\cite{Asimov_sig} respectively with all background contributions. The optimal cut defines two mutually exclusive regions that are used to probe for pomeron and photon processes separately:\\

\begin{itemize}
    \item \textbf{Photon-enriched region} --- events lying to the right of the optimal cut of  Figure~\ref{fig:score_distributions} (c).
    \item \textbf{Pomeron-enriched region} --- events lying to the left of the optimal cut of  Figure~\ref{fig:score_distributions} (c).
\end{itemize}

\vspace{0.2cm}

\noindent The domain separation displayed in Fig.~\ref{fig:score_distributions}(c) directly corresponds to the pomeron-induced vs non-peripheral $t\bar{t}X$ and photon-induced vs non-peripheral $t\bar{t}X$ classifiers shown in panels (a) and (b). Since events were uniquely indexed and no repetition was allowed across regions, this mapping is consistent and enables the subsequent extraction of the parameters $\mu_{\mathbb{P}}$ and $\mu_{\gamma}$ through the following statistical strategy. The two-step optimization yields two orthogonal event regions, defined such that no event contributes to more than one category. These regions are used as the sole binned inputs to profile–likelihood fit. Only per-bin Poisson constraints for MC‐statistical uncertainties are included via \texttt{autoMCStats} in \textsc{Combine}; no additional rate or shape systematics are applied.  For analysis category $c$ (e.g.\ photon-enriched or pomeron-enriched) and histogram bin $i$, the expected number of events is given by

\begin{equation}
\lambda_{c,i}(\mu_j,\boldsymbol\vartheta) =
\sum_{j \in \{\gamma,\mathbb{P}\}} \mu_j\,\vartheta^{S_j}_{c,i}\,T^{S_j}_{c,i}
+ \sum_{b\in B} \vartheta^{b}_{c,i}\,T^{b}_{c,i},
\end{equation}

\noindent where:

\begin{itemize}
    \item $\mu_j$ are the signal strength parameters, with $j=\gamma$ denoting the photon-induced process and $j=\mathbb{P}$ the pomeron-induced process. Each parameter rescales the corresponding SM.
    \item $T^{S_j}_{c,i}$ are the MC-predicted bin contents for the photon- or pomeron-induced signal templates in category $c$ and bin $i$, normalized to the SM.
    \item $B$ denotes the set of all considered background processes (e.g.\ non-peripheral $t\bar{t}X$ production).
    \item $T^{b}_{c,i}$ are the MC-predicted bin contents for background process $b$ in category $c$ and bin $i$.
    \item $\vartheta^{p}_{c,i}$ are the Barlow–Beeston nuisance parameters scaling the yield of process $p$ in bin $i$ of category $c$ to account for MC statistical fluctuations.
\end{itemize}

\begin{figure}[!ht]
  \centering
  \includegraphics[width=\columnwidth]{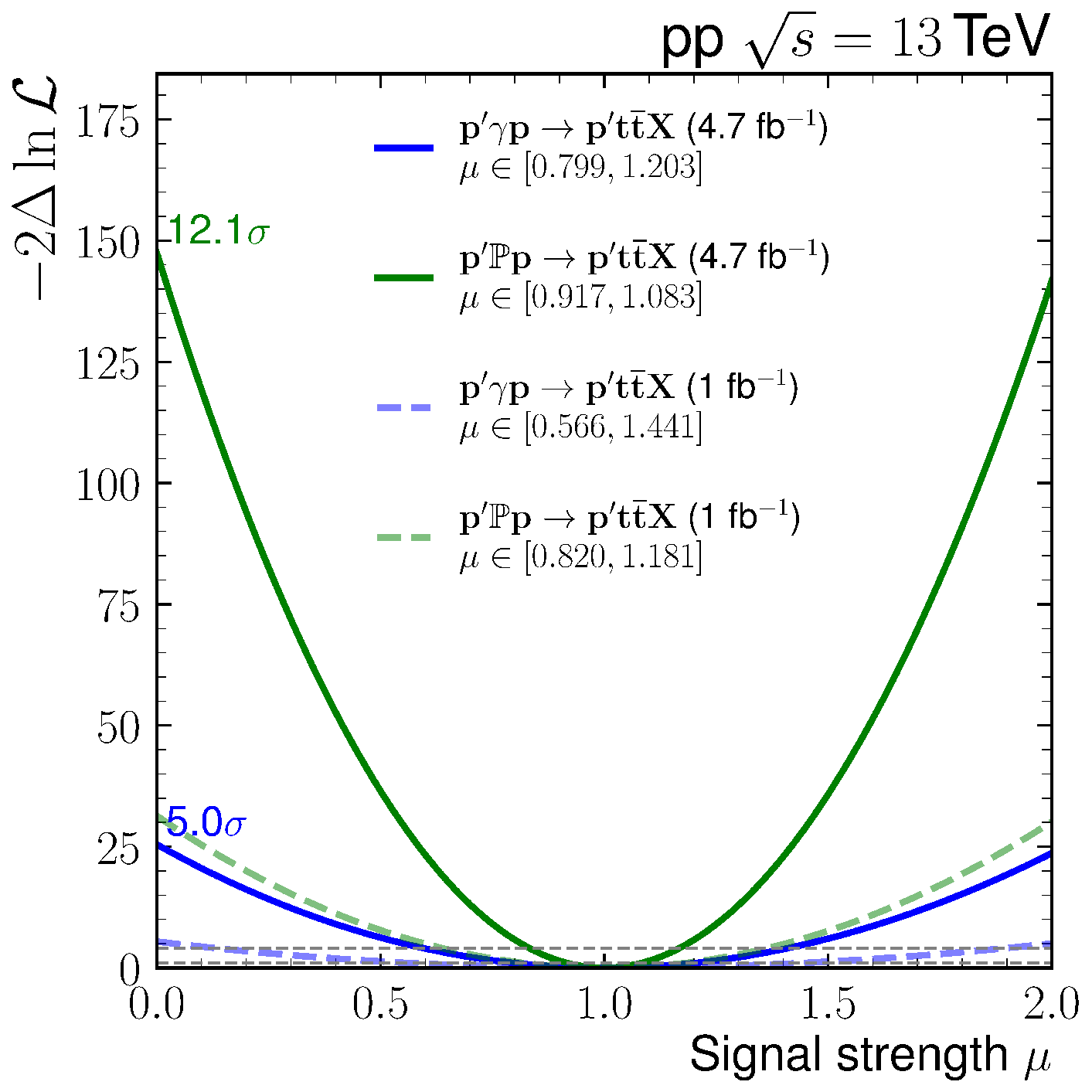}
  \caption{\justifying Profile‐likelihood scans at $68\%$ and $95\%$ CL for photon ($p'\gamma p \rightarrow p't\bar{t}X$) and pomeron ($p'\mathbb{P}p \rightarrow p't\bar{t}X$) signals, obtained from the Asimov–only fit. Here, $\mathcal{L}$ denotes the likelihood ratio used in the test statistic. The vertical axis value at $-2\Delta \ln{\mathcal{L}}= 25$ corresponds to the conventional $5\sigma$ discovery threshold.}
  \label{fig:likelihood_scan}
\end{figure}

\noindent The profile likelihood is constructed from the Asimov dataset:
\begin{equation}
L_{j}(\mu_{j},\boldsymbol\vartheta) =
\prod_{c,i} \mathrm{Pois}\!\left(n^{A}_{c,i} \;\middle|\; \lambda_{c,i}(\mu_{j},\boldsymbol\vartheta)\right)\,
C_{\mathrm{BB}}(\boldsymbol\vartheta),
\end{equation}
where $n^{A}_{c,i}$ are the expected bin counts in category $c$, bin $i$, and 
$C_{\mathrm{BB}}(\boldsymbol\vartheta)$ denotes the product of Poisson constraints implementing the Barlow–Beeston treatment of MC statistical uncertainties. The best-fit values, expected uncertainties, and one-dimensional likelihood scans for $\mu_\gamma$ and $\mu_{\mathbb{P}}$ are derived from this Asimov-only likelihood. These likelihood scans were performed using the \texttt{MultiDimFit} function of \textsc{Combine}. The one–dimensional profile–likelihood scans are shown in Fig.~\ref{fig:likelihood_scan}. For each curve, the parameter of interest is scanned while profiling the other signal strength and all Barlow–Beeston per–bin nuisance parameters. The reference at $-2\Delta \ln{\mathcal{L}}= 25$ corresponds to the $5\sigma$ discovery threshold. So this distribution indicates (green dashed line) that 1 fb$^{-1}$ of low pile-up $pp$ data would be enough to discover semi-exclusive pomeron-induced $t\bar{t}$.\\

\noindent Best–fit signal strengths for the set of alternative pomeron and photon schemes are compiled in Table~\ref{tab:mu_individuals}.
  
\begin{table}[!ht]
  \centering
  \resizebox{1.0\columnwidth}{!}{%

  \setlength{\tabcolsep}{6pt} 
  \footnotesize               
  \begin{tabular}{lc}
    \toprule
    \textbf{Photon scheme}         & $\mu \pm \mathrm{stat}$ \\
    \midrule
    MadGraph W.W.\ NLO             & $1.000^{+0.203}_{-0.201}$ \\
    PYTHIA8 Budnev LO              & $1.076^{+0.219}_{-0.216}$ \\
    PYTHIA8 Drees–Zeppenfeld LO    & $0.972^{+0.198}_{-0.195}$ \\
    \midrule
    \textbf{Pomeron scheme}        & $\mu \pm \mathrm{stat}$ \\
    \midrule
    Default                        & $1.000^{+0.083}_{-0.083}$ \\
    H1-A-NLO                       & $0.778^{+0.065}_{-0.065}$ \\
    H1-B-NLO                       & $1.052^{+0.088}_{-0.088}$ \\
    H1-B-LO                        & $0.933^{+0.078}_{-0.078}$ \\
    H1-AB-LO                       & $0.749^{+0.063}_{-0.062}$ \\
    \bottomrule
  \end{tabular}
  }
\caption{\label{tab:mu_individuals} 
Best‐fit signal strengths $\mu$ for photon and pomeron signals, with statistical 
uncertainties (68\% CL) from the Asimov–only fit at $4.7$ fb$^{-1}$.}
\end{table}

\vspace{-1.0cm}

\noindent Here, the yields are scaled to a scenario of 4.7fb$^{-1}$ luminosity that is sufficient to observe both semi-exclusive processes (continuous lines in Fig.~\ref{fig:likelihood_scan}) for a special $pp$ run with low pile-up. The quoted uncertainties are statistical at 68\% CL from the Asimov–only fit, no additional rate or shape systematics are profiled.\\

\noindent A summary for the final uncertainties is reported in Table~\ref{tab:mu_summary}. Here, the first uncertainty is statistical and the second uncertainty, labeled ``syst. scheme'' reflects the systematic uncertainties propagation obtained by repeating the Asimov procedure with the alternative photon and pomeron schemes described in section II and quoting the resulting envelope as a symmetric modeling component; background templates are not varied in this estimate.\\

\begin{table}[!ht]
  \centering
    \resizebox{1.0\columnwidth}{!}{%
  \begin{tabular}{@{}lc@{}}
    \toprule
    \textbf{Process}      \quad \quad  \quad \quad \quad \quad         & $\mu \pm \mathrm{stat}\pm \mathrm{syst. scheme}$ \\
    \midrule
    $p'\gamma p\to p't\bar{tX}$     & $1.000^{+0.203}_{-0.201}\,\text{(stat)}\,^{+0.076}_{-0.076}\,\text{(syst. scheme)}$ \\
    $p'\mathbb{P}p\to p't\bar{t}X$  & $1.000^{+0.083}_{-0.083}\,\text{(stat)}\,^{+0.251}_{-0.251}\,\text{(syst. scheme)}$ \\
    \bottomrule
  \end{tabular}
  }
    \caption{\label{tab:mu_summary}
    Best‐fit $\mu$ for photon and pomeron as signals with statistical (Asimov fit) 
    and modeling uncertainties (95\% CL).}
\end{table}

\noindent In this configuration, the pomeron–induced process $p'\mathbb{P}p\to p't\bar{t}X$ would reach an expected significance of $12.1\sigma$, corresponding to a clear observation in the Asimov model. The photon–induced process $p'\gamma p\to p't\bar{t}X$ would achieve an expected $5.0\sigma$, meeting the conventional discovery threshold. These projections indicate that, under realistic low pile‐up conditions, a dataset of order a few~$\mathrm{fb}^{-1}$ would suffice to establish both processes with high statistical precision, marking an important milestone for semi‐exclusive top‐quark pair production studies at the LHC.\\

\section{\label{sec:level6}Conclusions and Outlook}

\noindent Confirmation of photon or pomeron induced \(t\bar t\) production has not been reported at the LHC yet. These alternative mechanisms to the well understood \(q\bar q\) and gluon-gluon fusion and scattering channels set a stringent test to the SM due to the much lower predicted cross sections by 2 orders of magnitude. Semi-exclusive \(t\bar t\) production within $pp$ collisions with one intact proton was investigated from LHC to FCC energies by implementing photon flux models and pomeron data-driven flux and structure functions within MC simulation packages PYTHIA8 and MadGraph5\_aMC@NLO. Under this strategy the pomeron object is modeled non-perturbatively possessing a partonic structure with respective distributions that are estimated from deep inelastic scattering experiments.\\ 

\noindent Expected production cross-sections were found to have a systematic variation by scanning over all photon and pomeron scheme options by up to 62.5\% for pomeron-induced processes and 9.2\% for photon-induced events, systematic effect that is propagated to the final signal strength uncertainty predicted values. A marked energy evolution was established: from \(\sqrt s=13~\mathrm{TeV}\) to \(100~\mathrm{TeV}\), the semi-exclusive rates increased by factors of \(\sim50\) for \(\mathbb P p\) and \(\sim22\) for \(\gamma p\); the ratio \(\sigma_{\mathbb P}/\sigma_\gamma\) grew from \(8.6\) to nearly \(20\), consistent with the pomeron gluon fraction rising from \(\sim80\%\) toward \(\sim98\%\) in FCC kinematics.\\

\noindent Pomeron object was found to take energy from the source proton with a fraction of \(0<x_{\mathbb P}<0.79\). For the case of photon-induced events, \(x_\gamma\) concentrated at low values ($<$ 0.1) but extending over 0.5 translating into a total energy of $>$ 3.3 TeV leading to the appearance of forward gaps and low multiplicities.\\

\noindent CMS detector response was analyzed by interfacing the simulating events with Delphes fast detector simulation. At detector level, forward-activity observables were decisive; the minimum forward calorimetric energy within the HF acceptance emerged as the strongest single-variable separator against non-peripheral $t\bar{t}X$ background. Selected observables were used to train a neural network TensorFlow classifier built from forward-sensitive and global observables achieved \(\mathrm{AUC}_{\text{test}}=0.9917\pm0.0002\) for photon versus non-peripheral $t\bar{t}X$, enabling stable shape-based fits with strong generalization and with no signs of overfitting.\\

\noindent A profile-likelihood inference based on multivariate discriminants was implemented in order to obtain predictions for semi-exclusive $\mu$ signal strength sensitivity as would be measured by the CMS detector at CERN. Asimov-only profile-likelihood scans with Barlow–Beeston treatment of MC statistics were performed. In the context of low pileup pp data and by considering the statistical effects and the systematic contribution from photon/pomeron schemes to the total uncertainty, a 5-$\sigma$ significance is expected for pomeron-induced and photon-induced $t\bar{t}$ production modes with 1\,fb$^{-1}$ and 4.7\,fb$^{-1}$ integrated luminosity datasets respectively. This result comprises a systematic contribution to the total uncertainty from flux and structure functions of 75.1\% and 27.2\% respectively. At \(4.7~\mathrm{fb}^{-1}\) the resulting \(95\%\) CL intervals were \(\mu_\gamma\in[0.716,\,1.287]\) and \(\mu_{\mathbb P}\in[0.883,\,1.118]\). Further systematic effects usually present at the actual experimental measurement might increase the needed integrated luminosity for observation, though the aim of this work is to report on the particular systematic contribution from photon and pomeron schemes. It must be stressed that, beyond the quantified statistical precision, the diffractive predictions remain affected by large and essentially uncontrolled \emph{pomeron-model} uncertainties arising from the phenomenological treatment of the flux and DPDFs, reflecting the absence of a rigorous QCD factorization for hard diffractive $pp$ scattering. A more significant effect over the pomeron-induced channel with respect to the photon-induced by a factor of $\sim$3 is obtained. Which reflects the larger uncertainty over estimated pomeron structure in contrast with the photon case only requiring a photon flux function with no structure.\\ 

\noindent \noindent This result highlights the potential for experimental observation using Run 2 and Run 3 LHC data, allowing further studies within the forward physics program. In particular it is found that the systematic effect from photon and pomeron flux schemes and pomeron partonic densities is significant but would not be an obstacle for observation of semi-exclusive \(t\bar t\) before HL-LHC period. Looking ahead to the FCC regime, the pronounced energy scaling and the dominance of gluon-initiated diffractive dynamics make semi-exclusive \(t\bar t\) a promising laboratory for joint QCD--QED color-singlet studies and for precision tests of forward physics at the energy frontier. In addition, related topologies with one or two intact protons in photon and pomeron induced modes can be explored, further broadening the scope of forward-physics searches, for example, exclusive $\gamma\gamma$, $\gamma\mathbb{P}$, and $\mathbb{P}\mathbb{P}$ production, as well as semi-exclusive and exclusive single-top-quark ($tW$) production, can be investigated. The present work establishes a foundation for future studies by providing a phenomenological framework to estimate the luminosity required for a $5\sigma$ observation at the LHC and FCC. In contrast to previous analyses based on proton tagging and negligible background assumptions~\cite{Howarth:2020uaa}, the present approach incorporates realistic backgrounds and demonstrates that discovery-level separation of color-singlet topologies can be achieved without proton tagging, using only forward calorimetry through rapidity-gap and particle-multiplicity observables, while proton tagging remains a valuable complementary technique for future dedicated measurements on $t\bar{t}$ production rates.

\section*{Declarations}

\noindent \textbf{Funding} A.~Cota Rodríguez is supported by the Doctoral Program in Physics through a grant from the Government of Mexico, Secretaría de Ciencia, Humanidades, Tecnología e Innovación (\mbox{SECIHTI}).\\

\noindent \textbf{Competing Interests} The authors have no competing interests to declare that are relevant to the content of this article.\\

\noindent \textbf{Data Availability} This manuscript has no associated external data. The simulated datasets generated and/or analysed during the current study are not publicly available but are available from the corresponding author on reasonable request.\\

\noindent \textbf{Code Availability} This manuscript has no associated code/software. The code/software generated during and/or analysed during the current study is available from the corresponding author on reasonable request.

\section*{Acknowledgements}

The authors acknowledge the use of the YUCA high-performance computing resources at the Departamento de Matemáticas, Universidad de Sonora, for simulations and data analysis.


\bibliography{refs}

\end{document}